\documentclass[fleqn,usenatbib]{mnras}

\usepackage[normalem]{ulem}

\usepackage{graphicx}	
\usepackage{amsmath}	
\usepackage{amssymb}	

\input{vectors}


\usepackage[usenames,dvipsnames]{color}
\usepackage[]{color}

\newcommand{\arx}[1]{arXiv:1809.03666v1 [astro-ph.SR]}
\newcommand{\jphg}[1]{Journal of Physics G}

\title[Enhanced RGB mixing and Li-rich red-clump stars]
{Enhanced extra mixing in low-mass stars approaching the RGB tip and the problem of Li-rich red-clump stars}

\author[P. A. Denissenkov et al.]{Pavel A. Denissenkov$^{1,2,3\dagger}$\thanks{E-mail: pavelden@uvic.ca},
Simon Blouin$^{1,2}$, Falk Herwig$^{1,2,3\dagger}$, Jacob Stott$^{1}$ and \newauthor Paul R. Woodward$^{2,4}$
\\
$^{1}$Department of Physics \& Astronomy, University of Victoria, Victoria, BC V8W~2Y2, Canada\\
$^{2}$Joint Institute for Nuclear Astrophysics -- Center for the Evolution of the Elements (JINA--CEE), 
Michigan State University, \\ 640 South Shaw Lane, East Lansing, MI 48824, USA\\
$^{3}$CaNPAN, \href{https://canpan.ca}{https://canpan.ca}\\
$^{4}$LCSE and Department of Astronomy, University of Minnesota, Minneapolis, MN 55455, USA\\
$^\dagger$NuGrid Collaboration, \href{https://nugridstars.org}{https://nugridstars.org}\\}

\date{Accepted XXX. Received YYY; in original form ZZZ}

\begin{document}
\label{firstpage}
\pagerange{\pageref{firstpage}--\pageref{lastpage}}
\maketitle

\begin{abstract}
A few percent of red giants are enriched in Lithium with $A(\mathrm{Li}) > 1.5$.
Their evolutionary status has remained uncertain because these Li-rich giants
can be placed both on the red-giant
branch (RGB) near the bump luminosity and in the red clump (RC) region.
However, thanks to asteroseismology, it has been found that most
of them are actually RC stars. Starting at the bump luminosity,
RGB progenitors of the RC stars experience extra mixing in the radiative zone
separating the H-burning shell from the convective envelope followed by a series of
convective He-shell flashes at the RGB tip, known as the He-core flash.
The He-core flash was proposed to cause fast extra mixing
in the stars at the RGB tip that is needed for the Cameron-Fowler
mechanism to produce Li. We propose that the RGB stars are getting enriched
in Li by the RGB extra mixing 
that is getting enhanced and begins to produce
Li, instead of destroying it, when the stars are approaching the RGB tip. After a discussion of several mechanisms of the RGB
extra mixing, including the joint operation of rotation-driven meridional circulation and turbulent diffusion,
the Azimuthal Magneto-Rotational Instability (AMRI), thermohaline convection, buoyancy of magnetic flux tubes, and
internal gravity waves, and based on results of (magneto-) hydrodynamics simulations and asteroseismology
observations, we are inclined to conclude that it is the
mechanism of the AMRI or magnetically-enhanced thermohaline convection, that is most likely to support our hypothesis.
\end{abstract}



\section{Introduction}
\label{sec:intro}

\begin{figure}
  \centering
  \includegraphics[width=\columnwidth]{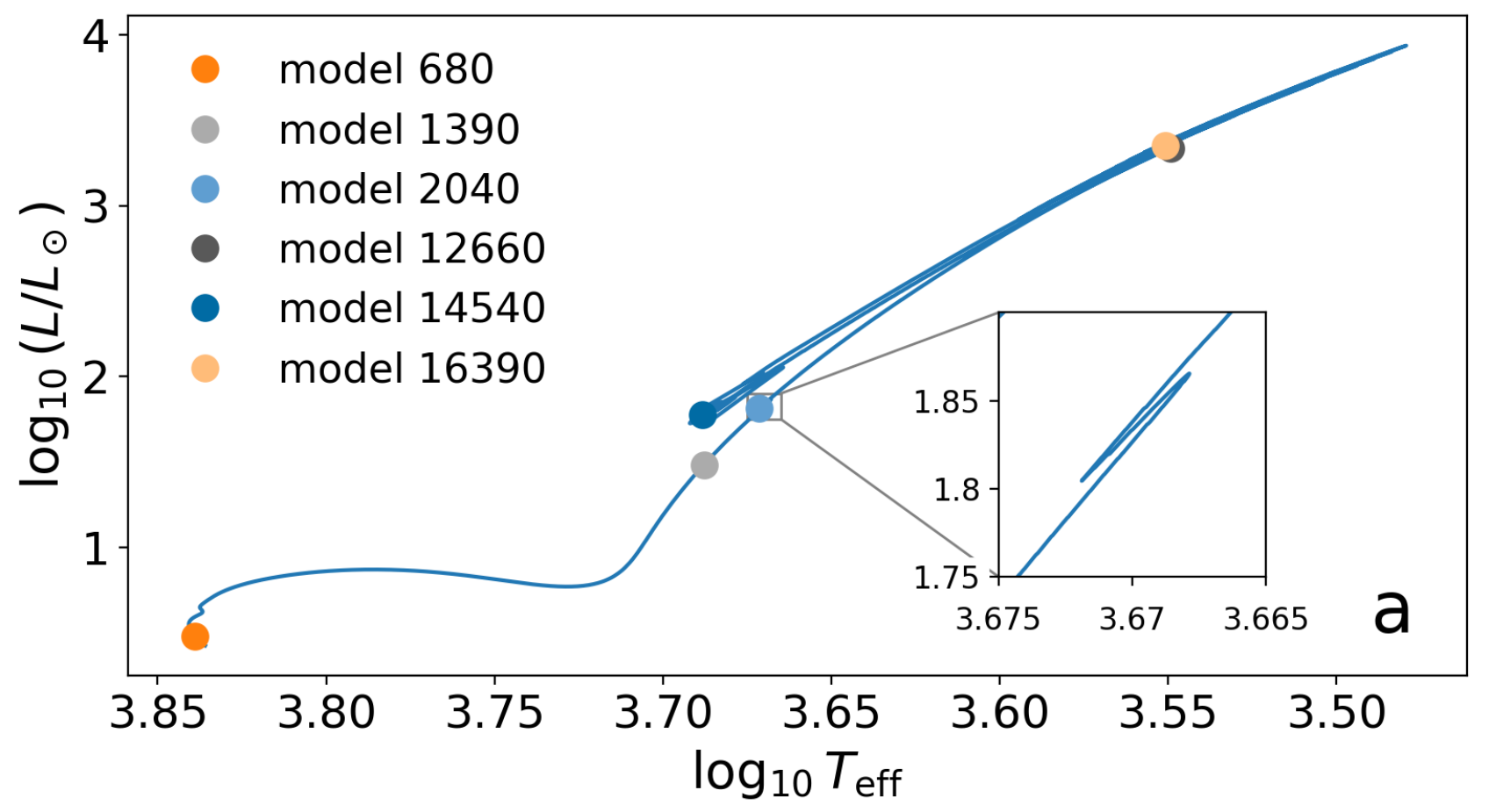}
  \includegraphics[width=\columnwidth]{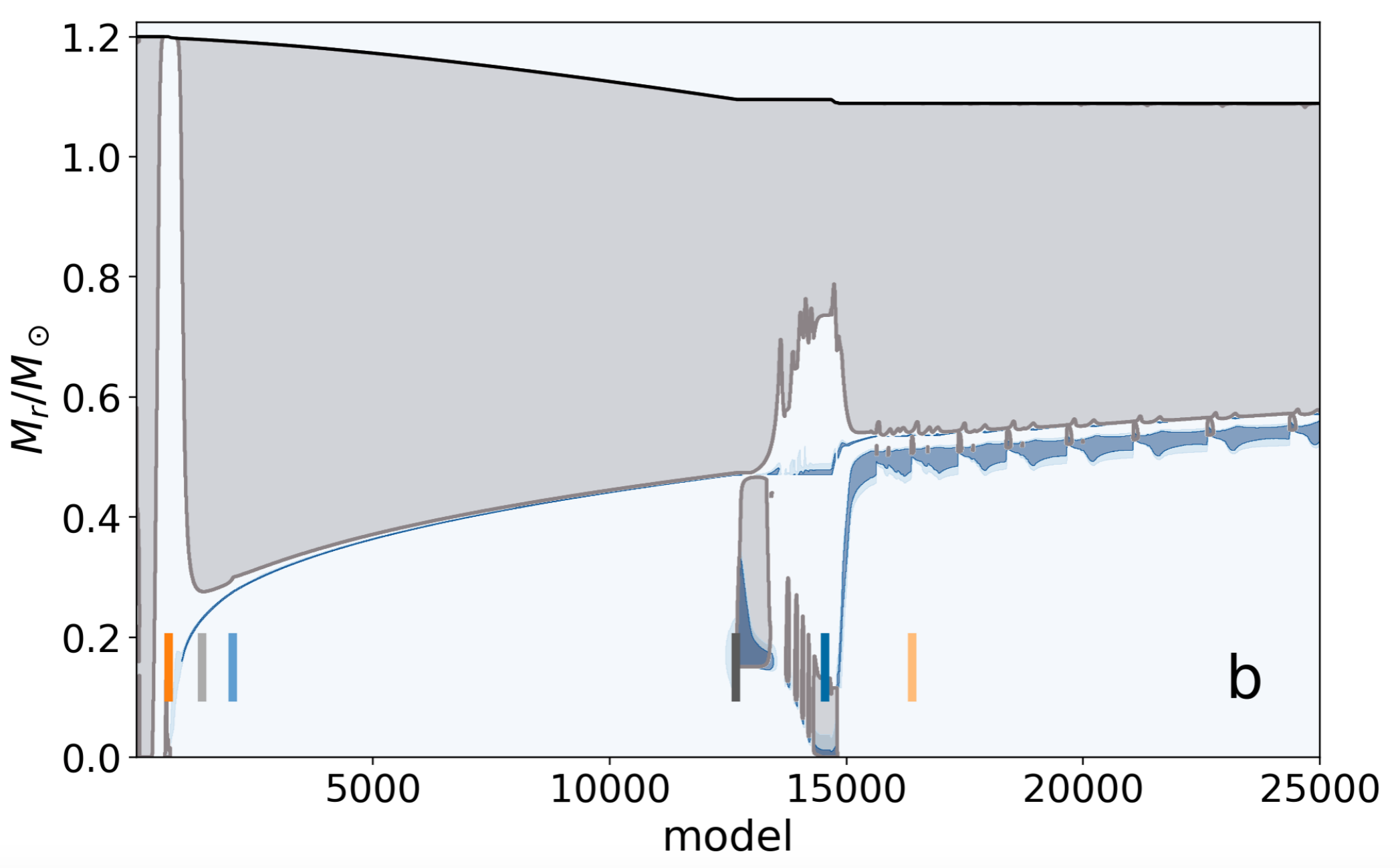}
  \caption{The track (panel a) and Kippenhahn diagram (panel b) for the evolution of a star
           with the initial mass $1.2\,M_\odot$ and metallicity [Fe/H]\,$=-0.3$ with model
           numbers (circles and vertical line segments of same colors) indicating its main phases, such as
           the main sequence (680), the end of the first dredge-up (1390), the bump
           luminosity on the red giant branch (2040), the first He-shell flash (12660), the He-core burning
           in the red-clump region of the horizontal branch (14540), and the beginning of the He thermal pulses
           on the asymptotic giant branch (16390). The gray-shaded regions in panel b are convective
           zones (envelope, cores and shells).
  }
  \label{fig:fig1}
\end{figure}

The standard stellar evolution theory predicts that the surface Li abundance has to decrease in a star with
a mass close to the solar one when it leaves the main sequence (MS) phase of H-core burning and begins to ascend
the red giant branch (RGB). This happens during the first dredge-up (FDU) episode, 
when the base of the deepening convective envelope of the star on the lower RGB reaches
the layers in which temperature was high enough, above 2.5 MK, to destroy Li on the MS, and now Li remaining
in the surface layers gets diluted by this convective mixing \citep{iben1967}. The FDU also reduces the surface
$^{12}$C to $^{13}$C isotopic and C to N elemental abundance ratios. At the end of the FDU, when
the base of the convective envelope stops deepening and begins retreating in front of the H-burning shell advancing in mass,
it leaves behind a small discontinuity in the H and other abundance profiles at the mass coordinate of
its deepest penetration. Later, when the H-burning shell crosses the discontinuity it
has to adapt to the slightly different chemical composition forcing the star to make a small zigzag on the
Hertzsprung-Russell diagram towards a lower luminosity and then to resume its RGB ascent (Figure \ref{fig:fig1}a). This temporarily
slows down its evolution and can be observed as a pile-up (a bump) of stars at the corresponding {\it bump luminosity}
in luminosity functions of populous globular clusters \citep{fusipecci1990}.

Observations show that above the bump luminosity, on the upper RGB, low-mass stars experience non-convective extra mixing
in their radiative zones separating the H-burning shell from the base of the convective envelope 
\citep[e.g.,][]{gilroy1991,gratton2000,smith2003,shetrone2019}
that further reduces their surface Li abundances along with the $^{12}$C/$^{13}$C and C/N ratios.
Later, at the tip of the RGB, the temperature in the He core becomes sufficiently high at some distance from the center to
ignite the triple-$\alpha$ reaction under electron-degenerate conditions. This {\it He-core flash} consists of a series of
convective He-shell burning events that gradually approach the center, lift the degeneracy and end up when He starts burning in a
non-degenerate convective core. By this time, the star arrives at the red-clump (RC) region of the horizontal branch (HB).
Finally, when He gets exhausted in the core, the star leaves the HB and begins to climb the asymptotic giant branch (AGB)
where it will experience intermittent H- and He-shell burning, the latter occuring in the form of thermal pulses.

The outlined standard scenario of the evolution of a low-mass star, which is illustrated by Figure~\ref{fig:fig1}, 
with observationally constrained mean rate and depth of
extra mixing on the upper RGB \citep{denissenkov2003} leads to a conclusion that the surface Li abundances in
low-mass RC stars have to be significantly reduced compared to their initial MS values of
$A(\mathrm{Li}) = \log_{10}[N(\mathrm{Li})/N(\mathrm{H})] + 12 = 3.3$ for the solar composition \citep{asplund2009}. This conclusion does not even
take into account the fact that the surface Li abundance in solar-type stars is observed to decline already on the MS, e.g.
by nearly 2 orders of magnitude in the Sun \citep{carlos2019}, as a result of another extra mixing of as yet unknown nature operating below
their convective envelopes \citep{richard2005,dumont2021}. Therefore, it was a surprise when \cite{kumar2020} discovered that all RC stars studied by them had
much higher Li abundances, by the factor of 40 on average, than those predicted by the standard
stellar evolution theory, and that the previously known population of a few percent of red giants with
$A(\mathrm{Li}) > 1.5$, the so-called {\it Li-rich giants}, all belonged
to an extended tail of the Li abundance distribution in RC stars.
That surprising indication of Li enrichment, ubiquitous in all low-mass stars, as claimed by \cite{kumar2020}, was challenged by \cite{chaname2022},
who showed that by allowing a realistic distribution of stellar masses and its corresponding Li abundances at the end of the MS,
the Li abundances of the bulk population of RC stars, those with $A(\mathrm{Li}) < 1.5$, are consistent with standard stellar models.  
Therefore, according to \cite{chaname2022}, the classical Li-rich giants with $A(\mathrm{Li}) > 1.5$ continue to be a problem for the standard stellar evolution theory, 
but there is no evidence, at least yet, that all the Li-normal RC stars with $A(\mathrm{Li}) < 1.5$ have also been enriched in Li.

Until about a decade ago, at least some of the Li-rich giants were thought to be RGB stars at the bump luminosity, because their luminosities and
effective temperatures did not allow to distinguish them from the RC stars. Their high Li abundances
used to be interpreted either as a consequence of their swallowing a giant planet or a brown dwarf with a preserved initial Li abundance
\citep{siess1999}
or as a result of enhanced RGB extra mixing with a rate significantly exceeding the one required for the explanation of
the observational evidence of the operation of RGB extra mixing in the majority of low-mass stars
\citep{charbonnel2000,denissenkov2000,denissenkov2004,denissenkov2012}. In the second case,
Li is synthesized via the mechanism proposed by \cite{cameron1971} in which $^7$Be produced in the vicinity of the H-burning shell
in the reaction $^3$He($\alpha,\gamma)^7$Be
has to be quickly transported by convection or extra mixing to colder layers where it will have enough time to capture electrons
to make $^7$Li, instead of being destroyed by proton captures. The identification of most of the Li-rich giants with RC stars was made
possible thanks to asteroseismology that demonstrated a clear separation of RGB and RC stars on the diagram displaying their
gravity-mode period spacing $\Delta\Pi_1$ versus large frequency separation $\Delta\nu$ constructed using photometric data
obtained with the space telescope {\it Kepler} \citep{kepler,bedding2011,mosser2014,kumar2020,deepak2021a,mallick2023}. Li abundances in these stars
were determined using ground-based stellar spectroscopy data from the LAMOST and GALAH surveys \citep{lamost,galah}.  
For reviews of Li-rich giants, readers are referred to \cite{martell2021} and \cite{yan2022}.

Recently, \cite{chaname2022} have questioned the conclusion that most of the Li-rich giants are low-mass RC stars. Their doubt is based on the fact
that the RC region can be occupied not only by low-mass stars with $M\la 2\,M_\odot$, that are indeed expected to undergo the He-core flash
preceded by extra mixing on the upper RGB, but also by more massive stars with $M\ga 2\,M_\odot$ that may only experience the FDU resulting in a moderate
decrease of their surface Li abundances. However, this controversy has soon been resolved by \cite{mallick2023} who show that whereas in their
smaller sample of 109 RC stars with $M\ga 2\,M_\odot$ the observed Li abundances do agree with predictions of the standard stellar
evolution theory, the presence of Li-rich giants, including a few super Li-rich ones with $A(\mathrm{Li}) \ge 3.3$ that were not explained
by \cite{chaname2022}, in their larger sample of 668 RC stars with $M\la 2\,M_\odot$ indicates that Li is indeed produced in their progenitors
somewhere between the end of RGB extra mixing and the RC phase. In this work, we put forward and substantiate a hypothesis that Li may be produced
in low-mass stars by the same RGB extra mixing of unknown nature yet that begins to manifest
itself at the bump luminosity and initially destroys Li but that is getting enhanced and starting to produce Li when the stars are approaching the RGB tip.

Most stellar evolution computations in this work, e.g. those results of which are shown in Figure~\ref{fig:fig1}, 
have been done for a star with the initial mass $1.2\,M_\odot$ and metallicity
[Fe/H]\,$=-0.3$\footnote{We use the standard stellar spectroscopy notation [A/B]\,$=\log_{10}[N_\star(\mathrm{A})/N_\star(\mathrm{B})] - 
\log_{10}[N_\odot (\mathrm{A})/N_\odot (\mathrm{B})]$, where $N_\star$ and $N_\odot$ are number densities of elements A and B in a star and the Sun.},
close to those of the majority of Li-rich RC stars studied by \cite{deepak2021}, using the MESA revision 7624 code \citep{paxton2011, paxton2013}
with the same input physics as described in Section 2 of \cite{denissenkov2017}, except that here we have included extra mixing
on the upper RGB and used the parameter $\eta_\mathrm{R} = 0.36$ in the Reimers formula for the RGB mass-loss rate \citep{reimers1975}.

Our paper is organized as follows.
In Section 2, we review various
physical mechanisms that were proposed to explain RGB extra mixing. 
Section 3 summarizes the previously proposed hypotheses of the Li enrichment of RC stars. 
In Section 4, we identify the RGB extra mixing mechanisms with diffusion coefficients that are expected to be rapidly increasing
with the luminosity and select the Azimuthal Magneto-Rotational Instability \citep[AMRI;][]{ruediger2014,ruediger2015}
as the most promising one to support our hypothesis. Then, we present and
discuss results of our computations of the evolution of a low-mass stellar model representing the Li-rich RC stars with RGB extra mixing getting
enhanced when it approaches the RGB tip. 
Section 5 concludes the paper with a brief discussion of our hypothesis and its supporting arguments.

\section{Mechanisms for the RGB extra mixing}

The radiative zones of low-mass stars, where the operation of extra mixing on the upper RGB is evidenced by the evolutionary
declines of the surface Li abundance, the isotopic $^{12}$C/$^{13}$C and the elemental C/N abundance
ratios in field, open and globular cluster stars, are convectively stable. Therefore, rising and sinking fluid elements
in these zones have to exchange heat by radiation diffusion with their surrounding stably stratified (with a positive entropy gradient) stellar
layers to be able to reduce the buoyancy force trying to keep them in place. This means that the corresponding rate of mixing, 
expressed as a diffusion coefficient $D_\mathrm{mix}  = f K$, should be proportional to the radiative diffusivity
\begin{equation}
K = \frac{4acT^3}{3\varkappa C_P\rho^2},
\label{eq:K}
\end{equation}
where $a$ is the radiation density constant, $c$ the speed of light in vacuum, $T$ the temperature, $\varkappa$
the Rosseland mean opacity, $C_P$ the specific heat at constant pressure, and $\rho$ the density. 
The factor $f$, which should be smaller than one, depends on physical parameters associated with a specific mixing mechanism.

\subsection{Rotationally-induced mixing}
\label{sec:rotmix}

\cite{sweigart1979} and \cite{smith1992} considered rotationally-induced meridional circulation as a mechanism for the RGB extra mixing. 
In fact, \cite{sweigart1979} were the first to conclude that RGB extra mixing
could begin to reach the H burning shell only after the latter had crossed and erased the chemical composition discontinuity left behind by
the base of the convective envelope at the end of the FDU. After that, above the bump luminosity, 
the mean molecular weight gradient $\nabla_\mu = (\partial\ln\mu /\partial\ln P)$ would remain zero
in the bulk of the radiative zone down to the vicinity of the H-burning shell where its increasing positive value serves as a barrier for
all types of extra mixing, thereby establishing their maximum depth.

\cite{zahn1992} conjectured that in rotating stellar radiative zones the meridional circulation had to compete in transporting angular momentum
with rotationally-induced turbulent diffusion. He assumed that the latter was much stronger in the horizontal than in vertical direction and
that, as a result, the radiative zones were in a state of shellular rotation with the angular velocity $\Omega$ depending only on the radius $r$. 
In the radiative zone of an upper RGB star
this competition is described by the following differential equation of the angular momentum transport
in Eulerian coordinates:
\begin{eqnarray}
\frac{\partial}{\partial t}\left(\rho r^2\Omega\right) = \frac{1}{5r^2}\frac{\partial}{\partial r}
\left[\rho r^4\Omega (U-5\dot{r})\right] \nonumber \\
+ \frac{1}{r^2}\frac{\partial}{\partial r} \left(\rho r^4\nu_\mathrm{v}\frac{\partial\Omega}{\partial r}\right),
\label{eq:amt}
\end{eqnarray}
where $\dot{r} = (\partial r/\partial t)_{M_r}$ is a rate with which a mass shell $M_r$ is approaching the H-burning shell,
$U$ is the radial component of the meridional circulation velocity, and
\begin{eqnarray}
\nu_\mathrm{v} = D_\mathrm{v} = \eta\frac{\left(r\frac{\partial\Omega}{\partial r}\right)^2}{N^2} K
\label{eq:Dv}
\end{eqnarray}
is the vertical component of both turbulent viscosity ($\nu_\mathrm{v}$) and turbulent diffusion coefficient ($D_\mathrm{v}$)
produced by the differential shellular rotation in a radiatve zone with $\nabla_\mu = 0$. 
In Equation~(\ref{eq:Dv}), $\eta\sim 0.01$\,--\,$0.1$ is a parameter, whose values in the indicated range were confirmed
by hydrodynamics simulations \citep{prat2013,prat2014,prat2016,garaud2017}, and
\begin{eqnarray}
N^2 = N_T^2 + N_\mu^2 = \frac{g}{H_P}\delta (\nabla_\mathrm{ad}-\nabla_\mathrm{rad}) + \frac{g}{H_P}\varphi\nabla_\mu
\label{eq:N2}
\end{eqnarray}
is the square of the Brunt-V\"{a}is\"{a}l\"{a} (buoyancy) frequency represented as a sum of its thermal ($N_T^2$) and
chemical composition ($N_\mu^2$) parts. In the expressions for the last two terms, $g$ is the local gravity, $H_P$ the pressure-scale height,
$\nabla_\mathrm{ad}$ and $\nabla_\mathrm{rad}$ the adiabatic and radiative temperature gradients,
logarithmic and with respect to pressure, while
$\delta = -(\partial\ln\rho/\partial\ln T)_{P,\mu}$ and $\varphi = (\partial\ln\rho/\partial\ln\mu)_{P,T}$
are determined by the equation of state.

At the same time, \cite{chaboyer1992} showed that the strong horizontal turbulence had to reduce the efficiency of radial mixing by
meridional circulation making it possible to describe it as a diffusion, rather than advection, process with the following coefficient:
\begin{eqnarray}
D_\mathrm{eff} = \frac{\left|rU\right|^2}{30D_\mathrm{h}},
\end{eqnarray}
where the coefficient of horizontal turbulent diffusion $D_\mathrm{h}$ was calculated using the radial and horizontal components of
the meridional circulation velocity, assuming that $D_\mathrm{h}\gg D_\mathrm{v}$ and $D_\mathrm{h}\gg |rU|$.

\citet[][hereafter DT00]{denissenkov2000b} solved Equation~(\ref{eq:amt}) with the prescriptions for the meridional circulation velocity and
turbulent diffusion from \cite{zahn1992} and their updates from \cite{maeder1996} and \cite{maeder1998} 
for the radiative zone of an upper RGB star and found that for reasonable surface
rotational velocities of RGB stars the combined diffusion coefficient $D_\mathrm{mix} = D_\mathrm{v} + D_\mathrm{eff}$, with
the first term dominating, provided a sufficiently fast rate for the RGB extra mixing. 
The magnitude of the circulation velocity calculated by DT00, $2\times 10^{-3}\ \mathrm{cm\,s}^{-1}$,
was very close to its value estimated by \cite{smith1992}, $3.6\times 10^{-3}\ \mathrm{cm\,s}^{-1}$,
they needed to reproduce the evolutionary decline of the carbon abundance 
on the upper RGB in the globular cluster M92 measured by \cite{carbon1982}.
Those results were obtained by DT00 under the assumption that the convective envelopes of RGB stars rotated
differentially keeping the specific angular momentum conserved through them. This assumption is supported by the results of 3D hydrodynamics
simulations of turbulent convection in the convective envelope of a low-mass RGB star reported by \cite{brun2009} and by the models of the evolution of
low-mass stars with rotation that explain the relatively fast rotational velocities of HB stars in the globular cluster M13 \citep{sills2000}.
The assumption of solid-body rotation of their convective envelopes would require unrealistically high surface rotational velocities for RGB stars. 
This conclusion agrees with the simulations of rotationally-induced mixing by
\cite{charbonnel2010} who used equations similar to those of \cite{zahn1992} but assumed solid-body rotation in convective envelopes
and, as a result, found that their rotationally-induced mixing on the upper RGB was too slow to explain the observational data. 

\cite{palacios2006} questioned the results of DT00 on the efficiency of rotationally-induced mixing in low-mass
stars on the upper RGB. They solved the equation of the angular momentum transport by the meridional circulation
and turbulent diffusion in Lagrangian coordinates for the entire evolution of a low-mass star from the MS to the upper RGB,
taking into account magnetic breaking of its envelope rotation on the MS and using different prescriptions for the coefficient of
turbulent diffusion. The angular velocity profiles in the radiative zones of the RGB bump luminosity models M2 and M6 displayed by \cite{palacios2006} 
in their Fig.~2 agree surprisingly well with the corresponding profile from Fig.~6 of DT00. Therefore, it is not clear why
the coefficient of vertical turbulent diffusion presented in Fig.~9 of \cite{palacios2006} for the model M4, which is the closest one to the model 
considered by DT00, is smaller by at least a factor of 10 than its counterpart presented in Fig.~5b of DT00. 
In any case, new data on internal rotation of low-mass subgiant and lower RGB stars obtained by asteroseismology have
revealed much flatter angular velocity profiles than those predicted by all 1D models that include only rotationally-induced
transport of angular momentum. 

\subsection{Angular momentum transport and mixing  driven by the azimuthal magneto-rotational instability (AMRI)}
\label{sec:amri}

\cite{beck2012} detected rotational splittings of mixed modes of solar-like oscillations in three lower RGB stars observed by the {\it Kepler} space telescope.
From an analysis of dipole modes they figured out that the cores of those stars rotated at least ten times faster than their envelopes
(for rotationally-induced RGB extra mixing that difference had to be nearly two orders of magnitude larger, e.g. see Fig. 6 of DT00), assuming that
the former and latter each rotated as a solid body. 
That conclusion was soon confirmed by \cite{mosser2012} who measured rotational splittings for a much larger sample of
low-mass red giants. \cite{deheuvels2014} added six low-mass subgiant stars to the sample of lower RGB stars studied by \cite{mosser2012} and showed
that their cores rotated by an order of magnitude faster than the cores of the RGB stars. They interpreted that as a signature of a more efficient 
than rotationally-induced transport of
angular momentum occurring in the latter. \cite{spada2016} modelled that transport as a diffusion process
and demonstrated that the observed evolutionary changes of the radiative core and convective envelope angular velocities 
$\Omega_\mathrm{core}$ and $\Omega_\mathrm{env}$ of 
the subgiant and RGB stars could be reproduced simultaneously with the angular momentum transport (AMT) diffusion coefficient
\begin{eqnarray}
D_\mathrm{AMT} = D_0\left(\frac{\Omega_\mathrm{core}}{\Omega_\mathrm{env}}\right)^\alpha,
\label{eq:DAMT}
\end{eqnarray}
where $D_0 \approx 1\ \mathrm{cm}^2\,\mathrm{s}^{-1}$ and $\alpha\approx 3$. They argued that such a power-law scaling 
with $\alpha\approx 2$\,--\,$3$ was
consistent with the dependence of a coefficient of turbulent viscosity on differential rotation obtained in numerical simulations of
the Azimuthal-Magneto-Rotational Instability (AMRI) by \cite{ruediger2015}. 
According to results reported by \cite{spada2016}, this AMT
begins to manifest itself approximately at the age when the H-burning shell is just established, i.e. long before 
the RGB star has reached the bump luminosity. An extrapolation of the $\alpha = 3$ curve from their Fig.~4
to a value of $\log g = 2.3$ at the bump luminosity of our $1.2 M_\odot$ model star
leads to an estimate of $D_\mathrm{AMT}\sim 10^6\ \mathrm{cm}^2\,\mathrm{s}^{-1}$ which is comparable to the value of
$D_\mathrm{mix}$ near the H-burning shell in Fig.~5d of DT00.
According to \cite{ruediger2014}, $D_\mathrm{AMT}/D_\mathrm{mix}  \propto \sqrt{\mathrm{Pm}}$, where $\mathrm{Pm}$ is
the magnetic Prandtl number representing a ratio of viscosity to magnetic diffusivity. 
Its value is between $0.01$ and $10$ in the radiative zone of low-mass RGB stars \citep{ruediger2015},
therefore $D_\mathrm{mix}\sim D_\mathrm{AMT}$ for the AMRI turbulence, unless a stable thermal stratification 
or a negative radial $\mu$ gradient prevents mixing.  
\cite{moyano2023} and \cite{dumont2023} have recently found additional observational support for the AMRI as the possible mechanism of AMT in
both low- and intermediate-mass stars on different evolutionary phases, starting at the subgiant branch (SGB) and including the HB. With their updated values of
$D_0 \approx 50\ \mathrm{cm}^2\,\mathrm{s}^{-1}$, $\alpha\approx 2$ \citep{moyano2023} and
$D_0 = 7.5\times 10^2 \ \mathrm{cm}^2\,\mathrm{s}^{-1}$, $\alpha\approx 1.5$ \citep{dumont2023}
for the low-mass stars the values of $D_\mathrm{AMT}\approx 5\times 10^5\ \mathrm{cm}^2\,\mathrm{s}^{-1}$ and
$D_\mathrm{AMT}\approx 7.5\times 10^5\ \mathrm{cm}^2\,\mathrm{s}^{-1}$ at the bump luminosity still remain sufficiently large
for the AMRI to be considered as a possible mechanism for the RGB extra mixing.

\subsection{Thermohaline mixing}
\label{sec:thm}

While doing 3D hydrodynamics simulations of convection triggered by the He-core flash in a $1\,M_\odot$ star at the RGB tip
\cite{dearborn2006} noticed some fluid motion outside the H-burning shell. In their follow-up paper
\cite{eggleton2006} found that a local inversion of the mean molecular weight profile $\mu(r)$ at the outer part of the H-burning shell
produced by the reaction $^3$He($^3$He,2p)$^4$He was driving that fluid motion, but they mistakenly attributed the cause of it to
the Rayleigh-Taylor instability. \cite{charbonnel2007} correctly interpreted that motion as thermohaline, or salt-fingering, convection driven by
a double-diffusive instability. It develops when diffusion of a destabilizing ingredient (salt in the ocean, nuclei contributing
to the difference in $\mu$ in the star) is less efficient than diffusion of heat that reduces the stabilizing effect of a difference 
in temperature between rising (and sinking) salt fingers and their surroundings. 
However, for a model of thermohaline convection to be able to reproduce
the observed evolutionary declines of the $^{12}$C/$^{13}$C and C/N ratios in upper RGB stars its salt fingers should have an aspect ratio of their
radial length to diameter $a\ga 7$. For the ideal gas equation of state, a simple linear analysis leads to the following expression
for the thermohaline diffusion coefficient:
\begin{eqnarray} 
D_\mathrm{th} = C_\mathrm{th}\frac{\nabla_\mu}{\nabla_\mathrm{rad}-\nabla_\mathrm{ad}}K,
\label{eq:Dth}
\end{eqnarray}
where $C_\mathrm{th} = 2\pi^2 a^2$ \citep{denissenkov2010} or $C_\mathrm{th} = (8/3)\pi^2 \alpha^2$ \citep{ulrich1972,charbonnel2007}, with
$\alpha$ also representing the salt-finger aspect ratio.
The MESA stellar evolution code that we have used in this work uses the parameterization 
$C_\mathrm{th} = (3/2)\alpha_\mathrm{th}$ referring to it as the ``Kippenhahn'' option motivated by the work of \cite{kippenhahn1980}.
The same observationally constrained value of $D_\mathrm{th}$ is obtained 
with $C_\mathrm{th}\approx 1000$ (for $a\approx 7$ or $\alpha\approx 6$), and $\alpha_\mathrm{th}\approx 667$.

\begin{figure}
  \centering
  \includegraphics[width=\columnwidth]{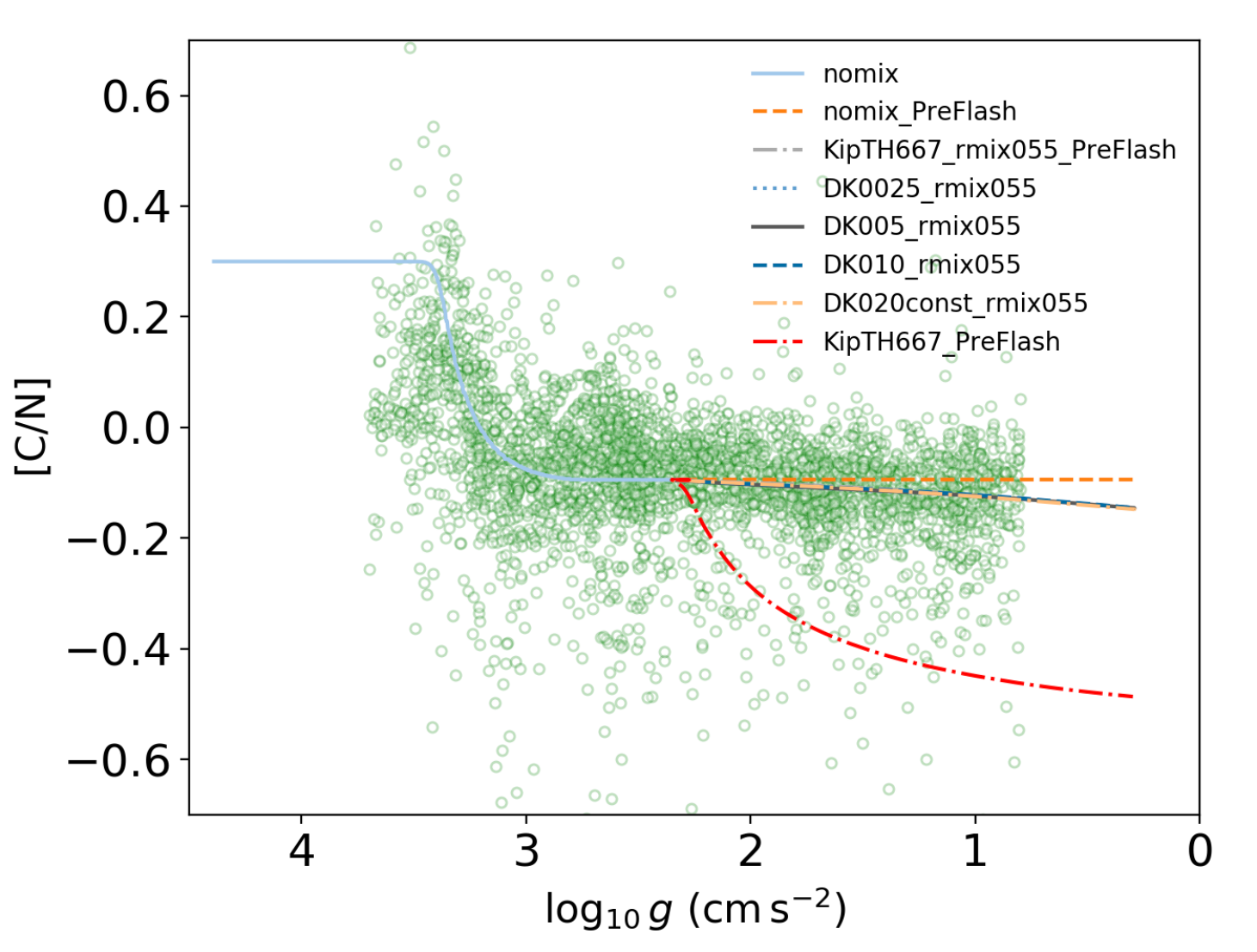}
  \caption{The evolutionary changes (the surface gravity $g$ decreases with time) of 
           the surface [C/N] elemental abundance ratio in our $1.2 M_\odot$ model star with $\mathrm{[Fe/H]}=-0.3$ between the MS and the RGB tip produced  
           by the first dredge-up (FDU) only (the solid blue curve followed by the dashed orange line) and by the RGB extra mixing  
           modelled as thermohaline convection using the MESA ``Kippenhahn'' option with the efficiency parameter $\alpha_\mathrm{th} = 667$
           (the dot-dashed red curve), as the same thermohaline convection but with the mixing depth fixed at $r_\mathrm{mix} = 0.055 R_\odot$
           (the dot-dashed gray line), and with the diffusion coefficient $D_\mathrm{mix} = f(L_\mathrm{bump})(L/L_\mathrm{bump})^{4/3}K$ for
           $f(L_\mathrm{bump})=0.0025$ (the dotted blue line), $f(L_\mathrm{bump})=0.005$ (the solid black line), and $f(L_\mathrm{bump})=0.010$
           (the dashed blue line), and $r_\mathrm{mix} = 0.055 R_\odot$. For comparison, we also show a model with the fixed $D_\mathrm{mix} = 0.02K$ and
           $r_\mathrm{mix} = 0.055 R_\odot$ (the dot-dashed orange line). The last five lines overlap in this Figure, but they 
           can all be seen separately in Figures \ref{fig:fig_c12c13} and  \ref{fig:fig_li_rc}. Green circles are the observational data from \protect\cite{shetrone2019}
           for field low-mass stars with $-0.35<\mathrm{[Fe/H]}<-0.25$. We had to shift our predicted [C/N] ratios assuming that their initial value is $+0.3$.
  }
  \label{fig:fig_cn}
\end{figure}

\begin{figure}
  \centering
  \includegraphics[width=\columnwidth]{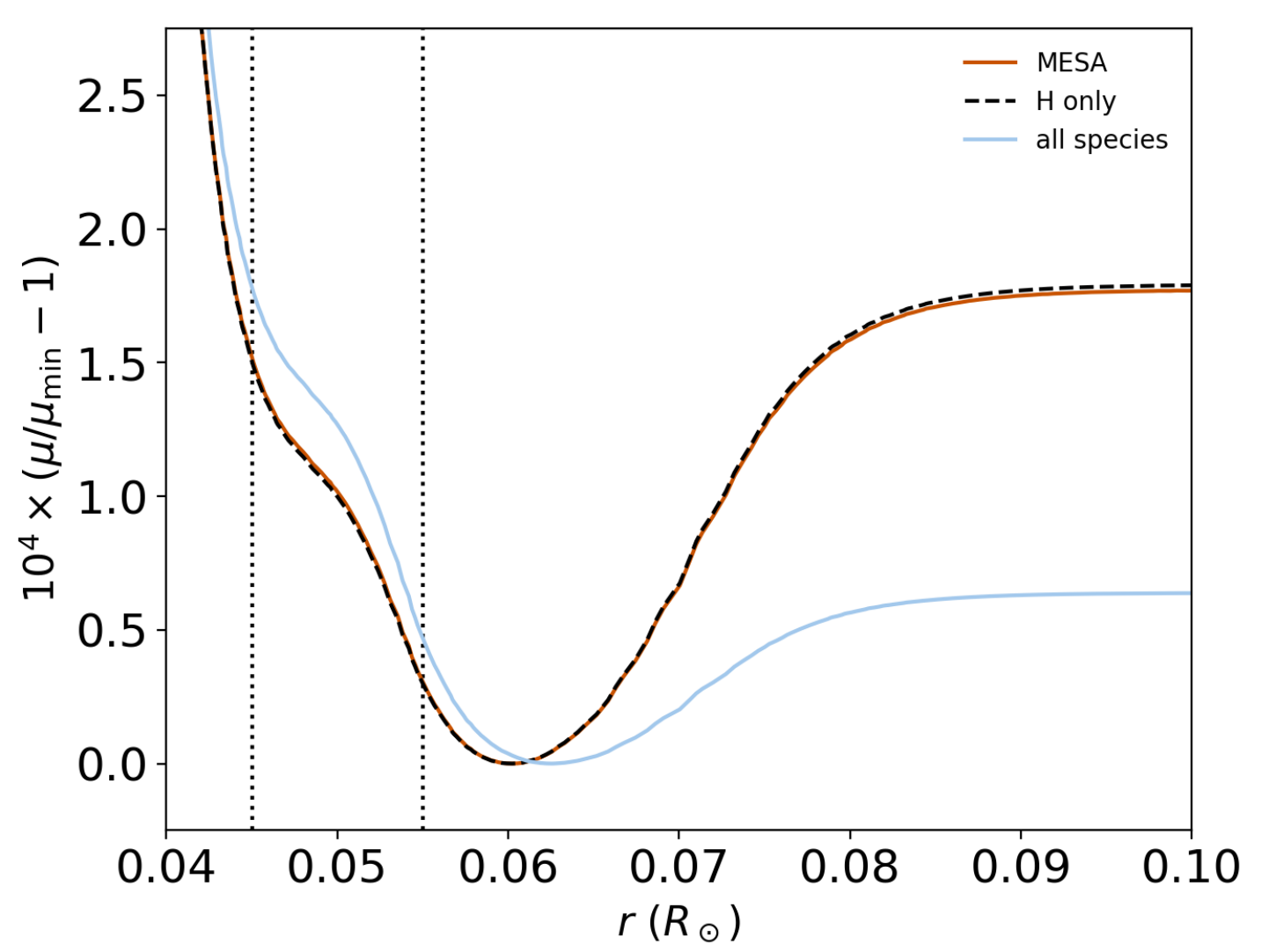}
  \caption{The stellar evolution code of MESA revision 7624 used in this work overestimates the depth of thermohaline mixing placing it at
           $r_\mathrm{mix}\approx 0.045\,R_\odot$ (the left vertical dotted line) because it uses for this the mean molecular weight
           based only on the H abundance. This results in too low [C/N] elemental abundance ratios predicted for the metallicity [Fe/H]=\,$-0.3$
           (Figure~\ref{fig:fig_cn}) compared to those observed in upper RGB stars with metallicities in the range
           $-0.4\le \mathrm{[Fe/H]}\le -0.2$ \protect\citep{shetrone2019}. Therefore, we fix the RGB extra mixing depth
           at $r_\mathrm{mix}\approx 0.055\,R_\odot$ (the right vertical dotted line) for all our models in this work.
  }
  \label{fig:fig_KipTH667_mu}
\end{figure}

When using the MESA ``Kippenhahn'' prescription for thermohaline mixing with $\alpha_\mathrm{th}\approx 667$ in our 
$1.2\,M_\odot$ model star we have obtained a too steep decline of the surface [C/N] ratio on the upper RGB,
with its total decrease produced by the FDU followed by RGB extra mixing between the bump luminosity and the RGB tip
$\Delta [\mathrm{C}/\mathrm{N}]\approx -0.8$ (dot-dashed red line in Figure~\ref{fig:fig_cn}).
This contradicts to a moderate change of $-0.3\la \Delta [\mathrm{C}/\mathrm{N}]\la -0.2$ in these mixing events observed
in low-mass RGB stars with metallicities in the range of $-0.4\le \mathrm{[Fe/H]}\le -0.2$ \citep[Fig. 3 and Table 2 in][]{shetrone2019}.
We have found that this discrepancy is caused by the MESA revision 7624 code overestimating the depth of thermohaline mixing because
it uses only the H abundance to calculate $\nabla_\mu$ in Equation (\ref{eq:Dth}). 
As illustrated by the left and right vertical dotted lines in Figure~\ref{fig:fig_KipTH667_mu},
the radius in the vicinity of the H-burning shell, below the local inversion of $\mu(r)$,  
at which the now increasing with depth $\mu$ approaches its value in the bulk of the radiative zone,
outside the H-burning shell, changes from $r_\mathrm{mix}\approx 0.045\,R_\odot$ to $r_\mathrm{mix}\approx 0.055\,R_\odot$
when $\mu$ is calculated using abundances of all available isotopes. Therefore, in this work we fix
the depth of RGB extra mixing in all our models at $r_\mathrm{mix}\approx 0.055\,R_\odot$, assuming that gas at $r<r_\mathrm{mix}$ with a higher $\mu$ than
in the bulk of the radiative envelope cannot rise up. 
A relatively small variation of [C/N] on the upper RGB obtained with the ``Kippenhahn'' prescription using this reduced mixing depth
is shown as a dot-dashed gray line in Figure~\ref{fig:fig_cn}.

We note that a relatively good quantitative agreement between the variation of the [C/N] ratio with the gravity seen in the APOGEE data
sampled by \cite{shetrone2019} for the Milky-Way field stars with metallicities $-0.35<\mathrm{[Fe/H]}<-0.25$ (green circles in Figure~\ref{fig:fig_cn})
and the decline of [C/N] predicted by our $1.2\,M_\odot$ model star with $\mathrm{[Fe/H]}=-0.3$ for the FDU (the solid blue curve) followed by 
its insignificant (because of the relatively high metallicity) additional decrease on the upper RGB modelled using different 
prescriptions for the standard and enhanced
extra mixing can be achieved only if we assume that the initial value of [C/N] is $+0.3$, or with the individually adjusted initial abundances of
$\mathrm{[C/Fe]}=+0.15$ and $\mathrm{[N/Fe]}=-0.15$ (all curves, except the solid blue, dashed orange, and dot-dashed red ones, in Figure~\ref{fig:fig_cn}).

\begin{figure}
  \centering
  \includegraphics[width=\columnwidth]{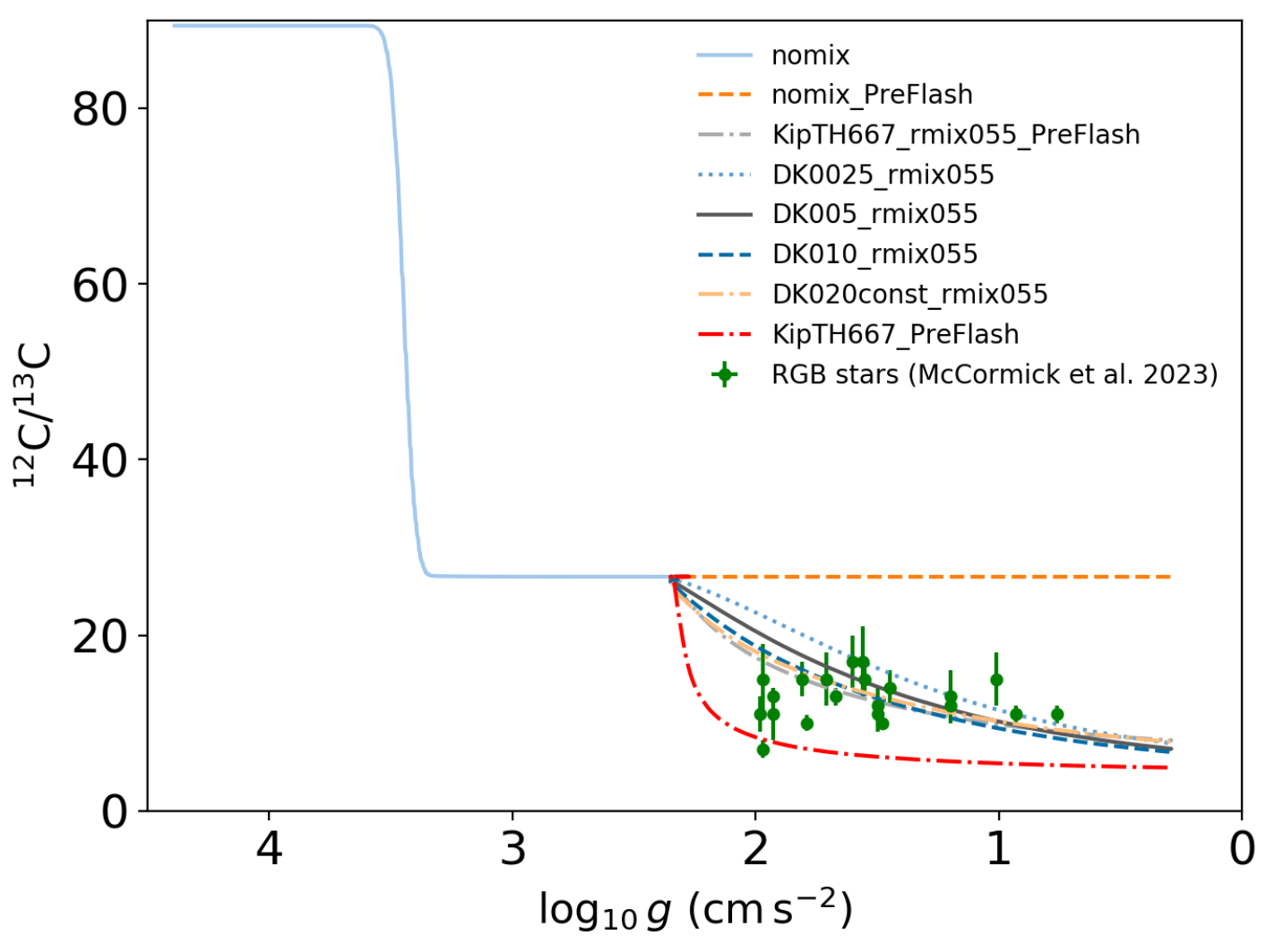}
  \caption{Same as in Figure \ref{fig:fig_cn}, but for the surface carbon isotopic ratio. Green circles with errorbars are APOGEE data for open-cluster
           RGB stars with initial masses lower than $2 M_\odot$, $\log_{10}g < 2$, and metallicities in the range $-0.36\la\mathrm{[Fe/H]}\la 0.28$ 
           from \protect\cite{mccormick2023}.
  }
  \label{fig:fig_c12c13}
\end{figure}

In lower-metallicity or in C-enhanced low-mass stars
the depth of the RGB thermohaline mixing has a different value, a smaller $r_\mathrm{mix}$ for a lower [Fe/H] 
and a larger $r_\mathrm{mix}$ for a C-enhanced mixture, but in all cases it remains in a range
between $r_\mathrm{mix}\approx 0.045\,R_\odot$ and $r_\mathrm{mix}\approx 0.06\,R_\odot$ \citep{denissenkov2008,denissenkov2009,denissenkov2010}.
We have checked that the same algorithm for the calculation of $\nabla_\mu$ is implemented in MESA at least up to the revision 15140, while
the most recent MESA revisions employ a different algorithm. Note that the problem with MESA overestimating the depth of RGB extra mixing
could also be revealed by comparing the observed and predicted carbon isotopic ratios
(Figure \ref{fig:fig_c12c13}).

Besides the technical difficulty in estimating a correct depth of the RGB thermohaline mixing, for which
some observational constraints still need to be used, a bigger problem is that
it needs a too large salt-finger aspect ratio of $a\ga 7$ to reproduce the observational data.
In the very low viscosity environment of stellar radiative zones, the rising and sinking salt fingers are subject
to the shear instability at their separating boundaries that should destroy their radially elongated structure. Such ``self destruction'' of
salt fingers was first predicted by \cite{kippenhahn1980} based on a simple analytical model and then it was
demonstrated in 2D and 3D numerical simulations by \cite{denissenkov2010} and \cite{denissenkov2011}, results of which were independently
confirmed by \cite{traxler2011}. In those simulations it was shown that the effective salt-finger aspect ratio is $\sim 0.5$,
which means that the efficiency of the RGB thermohaline mixing is $\sim 200$ times lower than what is required to explain
the observations. Therefore, at present we cannot consider thermohaline convection as a suitable model of RGB extra mixing. 

\subsection{Magnetically-assisted  mixing}

\subsubsection{Magnetic-buoyancy mixing}

One of such magnetic mixing mechanisms was proposed by \cite{busso2007} and developed by \cite{nucci2014}. 
It assumes that the radiative zones of upper RGB stars are filled with thin
donut-shaped magnetized flux tubes that are rising thanks to a difference in the densities inside and outside them caused by a magnetic pressure contribution
that makes them buoyant. However, the main drawback of those works was their omission to consider the relatively lengthy process of radiative heat exchange between
the rising flux tubes and their surroundings that was needed to maintain the difference in the densities. As a result of that omission,
the radial velocity of the flux tubes was overestimated by \cite{busso2007} by several orders of magnitude \citep{denissenkov2009}.
Indeed, for the initial values of the radius $a_0$ of the smaller circle of the donut-shaped flux tube and its radial velocity $v_0$
presented in Table~2 of \cite{busso2007} for the two RGB cases, and for the radiative diffusivity $K\sim 10^8 \mathrm{cm}^2\mathrm{s}^{-1}$
at the RGB mixing depth estimated from Figure~\ref{fig:fig_KipTH667} or taken from Table~1 of \cite{denissenkov2010},
we find that their corresponding ratios (P\'{e}clet numbers) of the thermal diffusion time $\sim a_0^2/K$ to
the advection time $\sim a_0/v_0$ per unit length of the larger circle of the flux tube are equal to $2184$ and $930$.
This means that the flux tubes with the parameters adopted by \cite{busso2007} rise too fast to be able to exchange heat with their surroundings,
which contradicts the assumption made by \cite{busso2007} that there is no difference in the temperature inside and outside the tubes.
Another drawback was that neither of those works discussed how the magnetized flux tubes
could be formed in the vicinity of the H-burning shell. \cite{denissenkov2009} suggested that such flux tubes were products of
the undular buoyancy instability \citep[][and references therein]{fan2001}, and that the azimuthal magnetic field of $\sim 100$ kG
needed for its development was generated by a strong differential rotation acting on and winding up a relatively weak ($\sim 10$ G) poloidal field.
\cite{denissenkov2009} also took into account heat exchange by radiative diffusion between rising flux tubes and their surroundings
and showed that it slowed down the magnetic-buoyancy mixing by nearly 5 orders of magnitude compared to the estimates of \cite{busso2007}.
At the same time, they found that a reduced mean molecular weight inside flux tubes formed in the region of its local inversion produced by the reaction
$^3$He($^3$He,2p)$^4$He compensated a substantial part of the negative effect of slow heat exchange. 
However, given that the magnetic-buoyancy model needs a rather strong differential
rotation to be present in the radiative zone, similar to that discussed in Section \ref{sec:rotmix}, which is not supported by the recent asteroseismology data,
we are forced to dismiss it as a plausible mechanism for RGB extra mixing.

\subsubsection{Thermohaline mixing in the presence of a radial magnetic field}
\label{sec:sec_mth}

\cite{harrington2019} invoked a radial magnetic field of the magnitude $B_r\sim 100$ G in their 3D direct numerical simulations of
thermohaline convection and showed that it could stabilize the growth of salt fingers
driven by the primary double-diffusive instability
against their destruction by the secondary shear instability in the radiative zone of an upper RGB star. That would increase the efficiency of
RGB thermohaline mixing by two orders of magnitude in agreement with the observations. 
However, more recently, \cite{fraser2024} have shown that the model of \cite{harrington2019}
does not accurately estimate the rate of thermohaline mixing for the values of model parameters appropriate for the radiative zone of an upper RGB star.
To resolve this issue, they have extended that model by including viscosity and magnetic resistivity, as well as temperature and composition
fluctuations, in their calculations of the growth of the parasitic shear instability. The extended model has recovered
the results of \cite{harrington2019} for the RGB star parameters 
and their used radial magnetic fields $100$ G and $1000$ G (see left panel in Figure 7 of \citealt{fraser2024}).
It is interesting and surprising that even much stronger radial magnetic fields, with magnitudes of the order of $100$ kG,
have recently been measured in the vicinity of the H-burning shell in 11 {\it Kepler} low-mass RGB stars using asteroseismology methods \citep{deheuvels2023}.
\cite{ligang2023} have detected similarly strong radial magnetic fields in another sample of 13 {\it Kepler} RGB stars with estimated masses
between $\sim 1 M_\odot$ and $\sim 1.5 M_\odot$. They have also found that neither the core rotation $\Omega_\mathrm{core}/2\pi\sim 1000$ nHz nor
the ratio $\Omega_\mathrm{core}/\Omega_\mathrm{env}\sim 10$\,--\,$100$ in these stars are different from those measured in other red giants.

\begin{figure}
  \centering
  \includegraphics[width=\columnwidth]{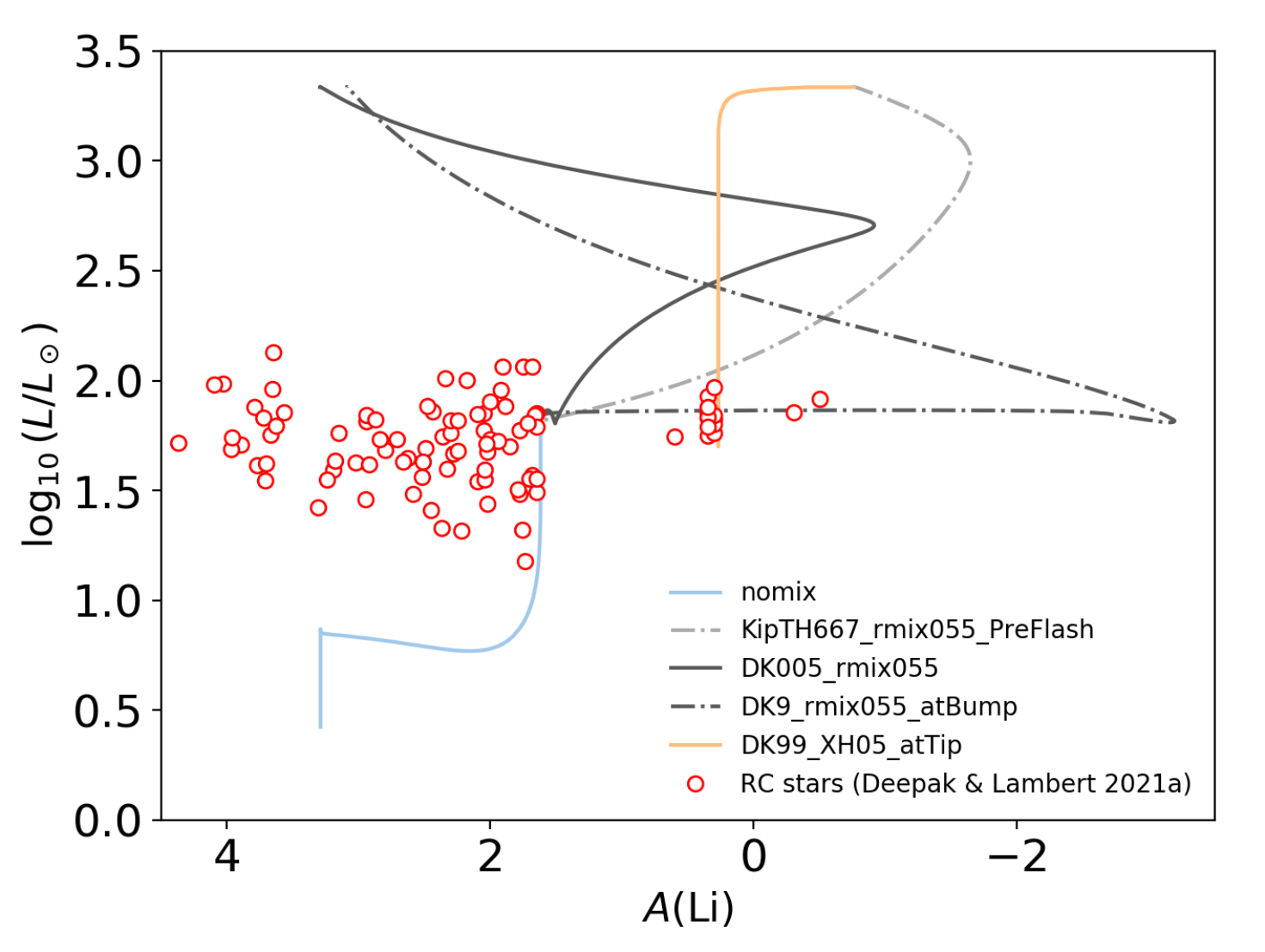}
  \caption{Li abundances for a subsample of RC stars with masses $0.77 M_\odot\leq  M\leq  1.96 M_\odot$ and metallicities [Fe/H]\,$\approx 0$
           from \protect\cite{deepak2021a} (red circles) are compared with Li abundances predicted by our 
           $1.2 M_\odot$ model star with $\mathrm{[Fe/H]}=-0.3$. The dot-dashed gray curve represents one of the same observationally-constrained cases of
           the standard RGB extra mixing, while the solid black curve corresponds to one of the same cases of
           enhanced extra mixing, as in Figures \ref{fig:fig_cn} and \ref{fig:fig_c12c13}.
           Two additional cases of enhanced RGB extra mixing were modelled, one with $D_\mathrm{mix} = 0.9K$ and $r_\mathrm{mix}=0.055 R_\odot$ 
           (the dot-dashed black curve), and the other with $D_\mathrm{mix} = 0.99K$ and $r_\mathrm{mix}$ placed inside the H-burning shell
           where the H mass fraction drops to $X=0.5$ (the orange curve). In the first case mixing starts at the bump luminosity and is described
           in Section \ref{sec:sec_mth}, and in the second case it occurs during the He-core flash and lasts as long as the He luminosity of 
           the first convective He shell exceeds $10^4 L_\odot$, as described in Section \ref{sec:sec_igwmix_he_core_flash}.
  }
  \label{fig:fig_li_mth_radigw}
\end{figure}

Our observationally-constrained constant diffusion coefficient for the standard RGB extra mixing has a value of $D_\mathrm{mix}\approx 0.02 K$
(dot-dashed orange curves in Figures \ref{fig:fig_cn} and \ref{fig:fig_c12c13}). 
According to \cite{fraser2024}, the salt-fingering convection can attain this mixing rate 
with $B_r\sim 100$\,--\,$1000$ G, and its magnetically-enhanced thermohaline diffusion coefficient is proportional to $B_r^2$.
If this is true and can be extrapolated to much stronger radial magnetic fields, 
then we can assume that in those 11+13 {\it Kepler} red giants with $B_r\sim 100$ kG the thermohaline diffusion coefficient
may approach its maximum possible value of $D_\mathrm{mix}\sim K$. The dot-dashed black curve in Figure \ref{fig:fig_li_mth_radigw} shows
that such enhanced RGB extra mixing could also lead to a substantial enrichment of RC stars in Li, provided it started
at the bump luminosity and the radial magnetic field of this strength was maintained in the radiative zone during the entire
RGB evolution. That would be an alternative to the AMRI as the essential element of our hypothesis (Section \ref{sec:our_hypo}), 
in which the observed spread of Li
abundances in RC stars could be explained by a range of radial magnetic field magnitudes in RGB stars.
Of course, we still have to understand the origin of these magnetic fields in radiative zones of RGB stars
\citep[e.g., see Section 4.4 in][]{ligang2023}.

\subsection{Mixing by internal-gravity waves}
\label{sec:IGWs}

Internal gravity waves (IGWs) or g modes of stellar oscillations are stochastically excited by turbulent fluid motion at a convective-zone boundary
and some of them can propagate through an adjacent radiative zone \citep{press1981}. Following the work of \cite{garcialopez1991},
\cite{denissenkov2003b} proposed that IGWs could produce partial mixing of a He- and C-rich radiative zone below the convective envelope
in low-mass AGB stars. It could gently inject protons into that zone, where the reactions $^{12}$C(p,$\gamma)^{13}$N and $^{13}$N(e$^+\nu)^{13}$C
would then form a sufficiently wide $^{13}$C pocket necessary for the main slow ({\it s}) neutron-capture process
to occur there under radiative conditions between He-shell thermal pulses \citep{straniero1995}.

\begin{figure}
  \centering
  \includegraphics[width=\columnwidth]{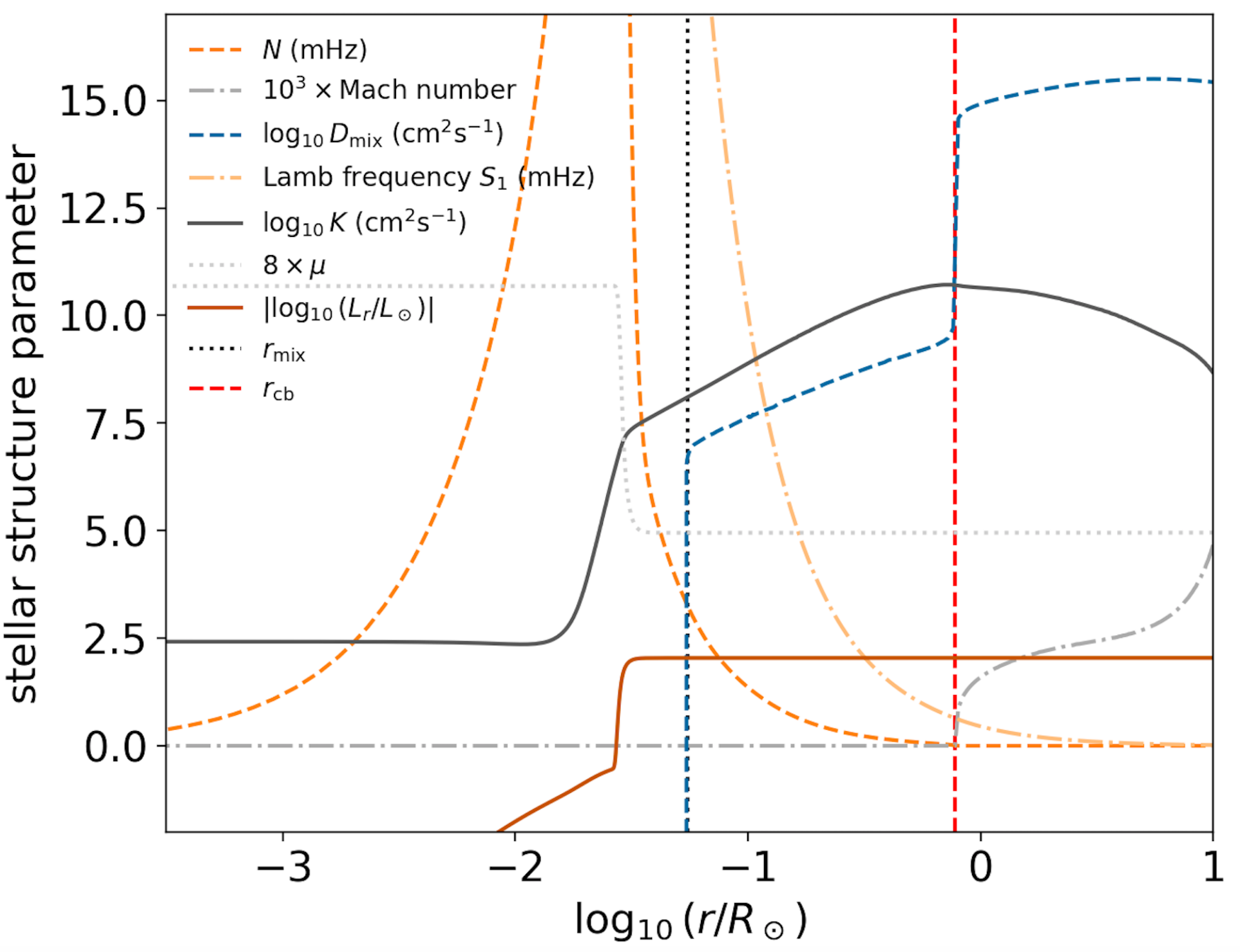}
  \caption{A snapshot of the internal structure of our $1.2 M_\odot$ model star, immediately above the bump luminosity, that includes
           an outer part of its He electron-degenerate core (where the mean molecular weight $\mu$ has reached its maximum value and stopped changing) 
           and an inner part of its convective envelope (above the vertical dashed red line representing the convective boundary).
           The radial profiles of the stellar structure parameters relevant to the different mechanisms of RGB extra mixing discussed in
           text are plotted, including the buoynacy frequency $N$ and the radiative diffusivity $K$. In the radiative zone, the diffusion coeffient $D_\mathrm{mix}$
           is calculated using the MESA ``Kippenhahn'' option with the efficiency parameter $\alpha_\mathrm{th} = 667$
           (the dashed blue curve at $r<r_\mathrm{cb}$) but with the mixing depth fixed at $r_\mathrm{mix} = 0.055 R_\odot$,
           as guided by Figure~\ref{fig:fig_KipTH667_mu}.
  }
  \label{fig:fig_KipTH667}
\end{figure}

For a g mode with the angular frequency $\omega$ and degree $l$, the horizontal wave number is
$k_\mathrm{h} = \sqrt{l(l+1)}/r$, while the vertical wave number $k_\mathrm{v}$ can be determined from the well-known dispersion relation for IGWs
\begin{eqnarray}
\frac{\omega^2}{N^2} = \cos^2\theta = \frac{k_\mathrm{h}^2}{k_\mathrm{h}^2+k_\mathrm{v}^2},
\label{eq:igw_disp}
\end{eqnarray}
where $N$ is the Brunt-V\"{a}is\"{a}l\"{a} (buoyancy) frequency from Equation (\ref{eq:N2}), and $\theta$ is the angle between the vertical (radial) direction and 
a plane of constant phase, the latter being parallel to directions of both wave's fluid oscillation and group velocity.
IGWs can propagate only in a region where $\omega < N$ and $\omega < S_l$, where $S_l = k_\mathrm{h}c_s = \sqrt{l(l+1)}c_s/r$ is
the Lamb frequency for a p mode (sound wave) with the angular degree $l$. 
For our $1.2\,M_\odot$ model star the radial profiles of $N$ and $S_1$ are plotted in Figures~\ref{fig:fig_KipTH667} and \ref{fig:fig_flash_rgb}
for the evolutionary stages immediately above the bump luminosity and at the beginning of the He-core flash, respectively.
For progressive and standing IGWs with
$\omega\ll N$ their transverse nature and Equation (\ref{eq:igw_disp}) lead to the following relation between their
vertical and horizontal velocity components:
\begin{eqnarray}
\frac{u_\mathrm{v}}{u_\mathrm{h}} = 
\frac{k_\mathrm{h}}{k_\mathrm{v}} \approx \frac{\omega}{N} \ll 1.
\label{eq:uvuh}
\end{eqnarray}
Hence, fluid in such waves oscillates in nearly horizontal directions, thus
producing a horizontal velocity shear. Like in the case of differential shellular rotation that we discussed in Section
\ref{sec:rotmix}, in the presence of a strong radiative heat diffusion this shear may drive small-scale turbulent mixing
in the radial direction. Using Equation (\ref{eq:Dv}), its diffusion coefficient can be estimated as
\begin{eqnarray}
D_\mathrm{IGW} =  \eta\frac{(k_\mathrm{v}u_\mathrm{h})^2}{N^2}K \approx \eta\frac{(\nabla\times\mathbf{u})^2}{N^2}K.
\label{eq:digw}
\end{eqnarray}

The second mechanism by which IGWs may produce some mixing was discussed in detail by \cite{schatzman1996}, who 
considered it as a result of a random walk of tracer particles being pushed by an ensemble of g-mode oscillations
with different wavelengths, frequencies, and amplitudes. In the ideal case of adiabatic oscillations, the root-mean square displacements of
the tracer particles are all equal to zero, but in the presence of heat exchange between oscillating g modes and their surroundings
by radiative diffusion this is not necessarily true. An analytical prescription for the diffusion coefficient associated with
this mechanism of IGW mixing was developed and implemented
to interpret the Li and Be depletions in the Sun and other low-mass MS stars of different ages by \cite{montalban1994} and \cite{montalban2000},
and it was also used to model the formation of the $^{13}$C pocket in low-mass AGB stars by \cite{denissenkov2003b}.
The weakest part of such analytical prescriptions, including the one based on Equation (\ref{eq:digw}), is that
their estimated final diffusion coefficients strongly depend on power (or velocity) spectra of IGWs generated at the convective-zone
boundary and on their attenuation in the adjacent radiative zone through which they propagate.

Until recently, only simple semi-empirical and analytical approaches have been employed to model IGW spectra and attenuation.  
This work has motivated us to perform the first 3D hydrodynamics simulations of turbulent convection and 
IGWs in a $4\pi$ sphere encompassing the lower part of the convective
envelope and a substantial part of the radiative zone of our $1.2\,M_\odot$ upper RGB model star.
Results of these simulations with nominal heating and realistic opacities for the RGB-tip model
are presented elsewhere \citep{blouin2023}, while here we summarize only
the most important of them that are relevant for the present work. First, the mean vorticity of fluid motion in radiative layers
at about one pressure scale height below the convective envelope scales with the total luminosity as
$|\nabla\times\mathbf{u}|\propto L^{1/4}$, rather than obeying our anticipated scaling law 
$|\nabla\times\mathbf{u}|\propto L^{2/3}$ (see Section~\ref{sec:igw_enhanced_mix}).
Second, when the vorticities measured in the radiative zone of the RGB-tip model from its
highest-resolution 3D hydrodynamics simulation are substituted in Equation (\ref{eq:digw}) with a probably upper-limit estimate of 
$\eta = 0.1$ the resulting values of $D_\mathrm{IGW}$ are at least 1 order of magnitude smaller and they decrease
with a depth in the radiative zone much faster than what is needed to
reproduce the observational data. Third, a diffusion coefficient in the radiative zone measured directly using
the tracer fluid Gaussians has even smaller values. Therefore, at present it seems highly unlikely that the RGB extra mixing
is produced by IGWs. One caveat of our hydrodynamics simulations is that they only include a small portion of
the convective envelope, and simulations with a larger portion of the convective envelope yield higher IGW velocities.

\section{Previous hypotheses proposed to explain the Lithium enrichment of RC stars}

In Figure~\ref{fig:fig_li_rc} we compare the evolution of the surface Li abundance in our $1.2\,M_\odot$ model star between the bump
luminosity and the RGB tip or its arrival at the RC region predicted using different prescriptions for extra mixing on the upper RGB 
and during the He-core flash with
the Li abundances in RC stars having masses in the range $0.77\le M/M_\odot\le 1.96$ and metallicities close to [Fe/H]\,$=0$
taken from \cite{deepak2021a}. The dashed orange and dot-dashed gray lines in this figure show that neither the evolution
with no extra mixing nor the one with thermohaline mixing on the upper RGB can reproduce the observed Li enrichment of RC stars.
In this Section we briefly discuss previous hypotheses that were put forward to explain this discrepancy, while
our own alternative explanation of it will be presented in the next Section.

\subsection{Lithium enrichment by IGW mixing triggered by the He-core flash} 
\label{sec:sec_igwmix_he_core_flash}

\begin{figure}
  \centering
  \includegraphics[width=\columnwidth]{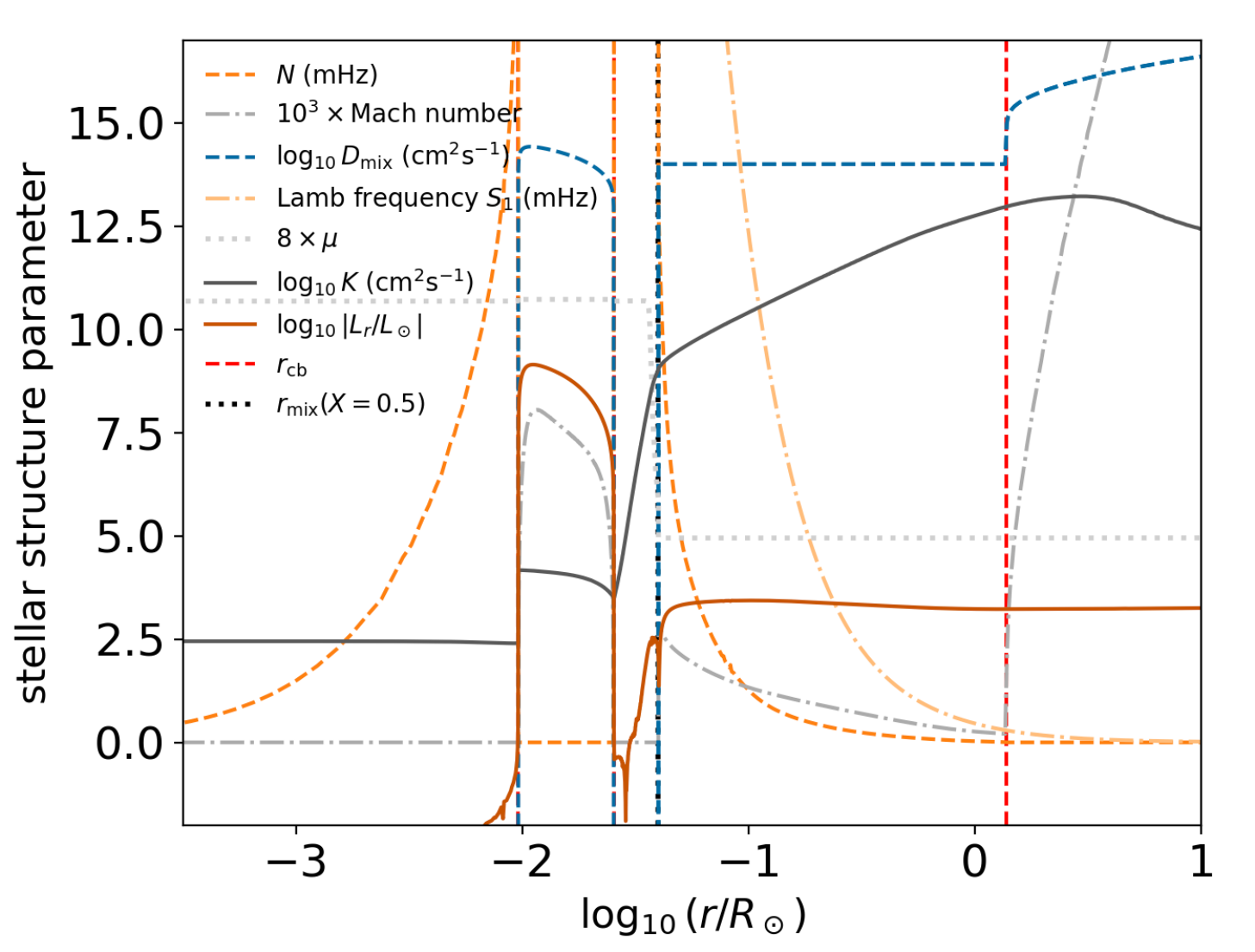}
  \caption{Similar to Figure \ref{fig:fig_KipTH667}, but for the RGB-tip model with the first convective He-shell zone developing and
           with IGW mixing in the radiative zone modelled similar to \protect\cite{schwab2020}, as explained in caption to Figure \ref{fig:fig_li_rc}.
  }
  \label{fig:fig_flash_rgb}
\end{figure}

\begin{figure}
  \centering
  \includegraphics[width=\columnwidth]{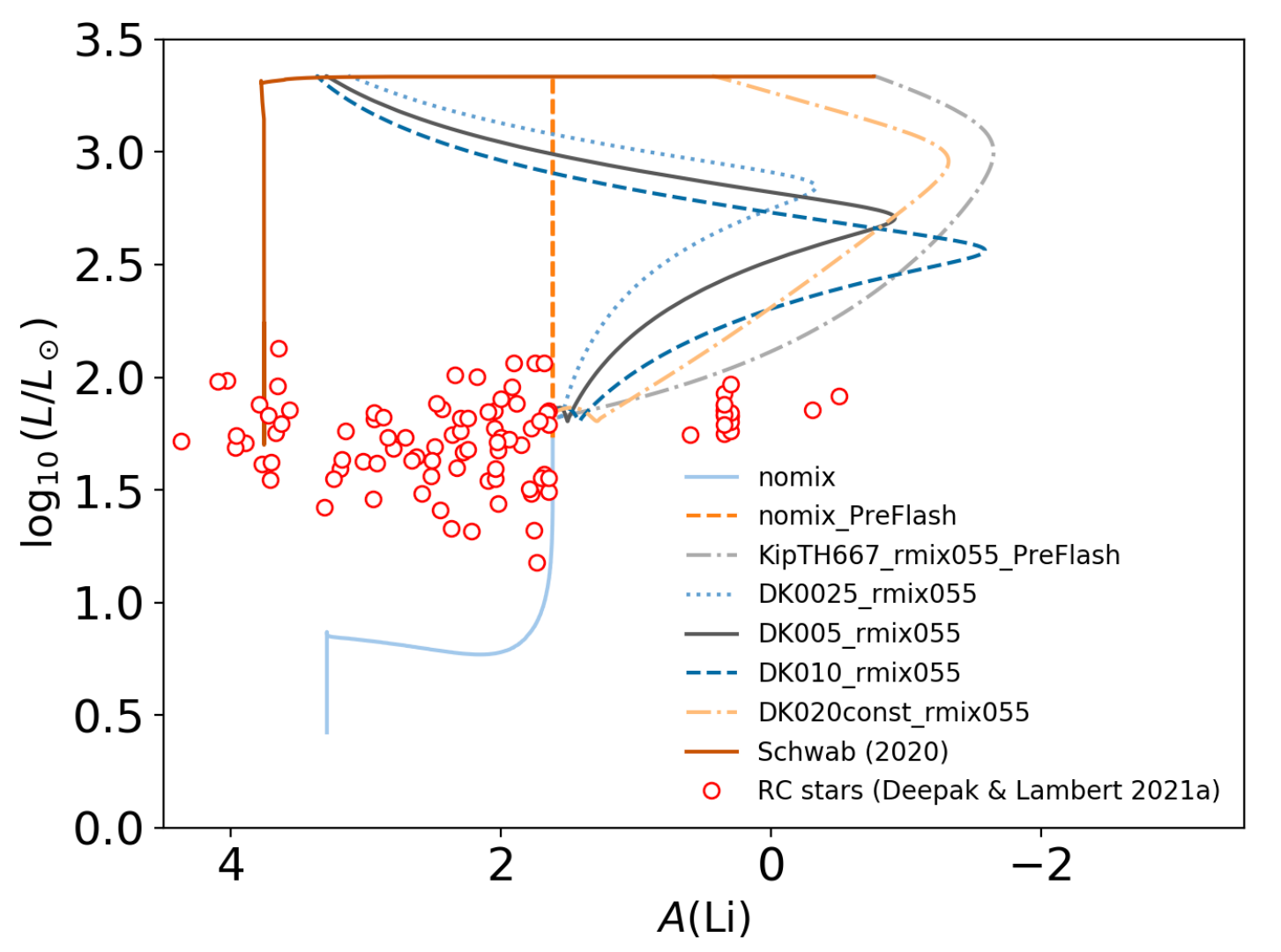}
  \caption{Similar to Figure \ref{fig:fig_cn}, but for the surface Li abundance and with the surface gravity replaced by
           the luminosity that increases with time. Additionally, the solid brown line represents
           the results of our attempt to repeat the IGW mixing calculations of \protect\cite{schwab2020}. We used his diffusion coefficient
           $D_\mathrm{mix} = 10^{14}\,\mathrm{cm}^2\,\mathrm{s}^{-1}$ in the radiative zone of our $1.2 M_\odot$ model star where
           the H mass fraction was $X>0.5$ and kept it constant while the luminosity of the first convective He-shell flash remained
           above $10^4 L_\odot$ during the He-core flash at the RGB tip. 
           Our predicted Li abundances are compared with those measured by \protect\cite{deepak2021a} in RC stars that
           have masses between $0.77 M_\odot$ and $1.96 M_\odot$ and metallicities close to [Fe/H]\,$=0$.
  }
  \label{fig:fig_li_rc}
\end{figure}

\begin{figure}
  \centering
  \includegraphics[width=\columnwidth]{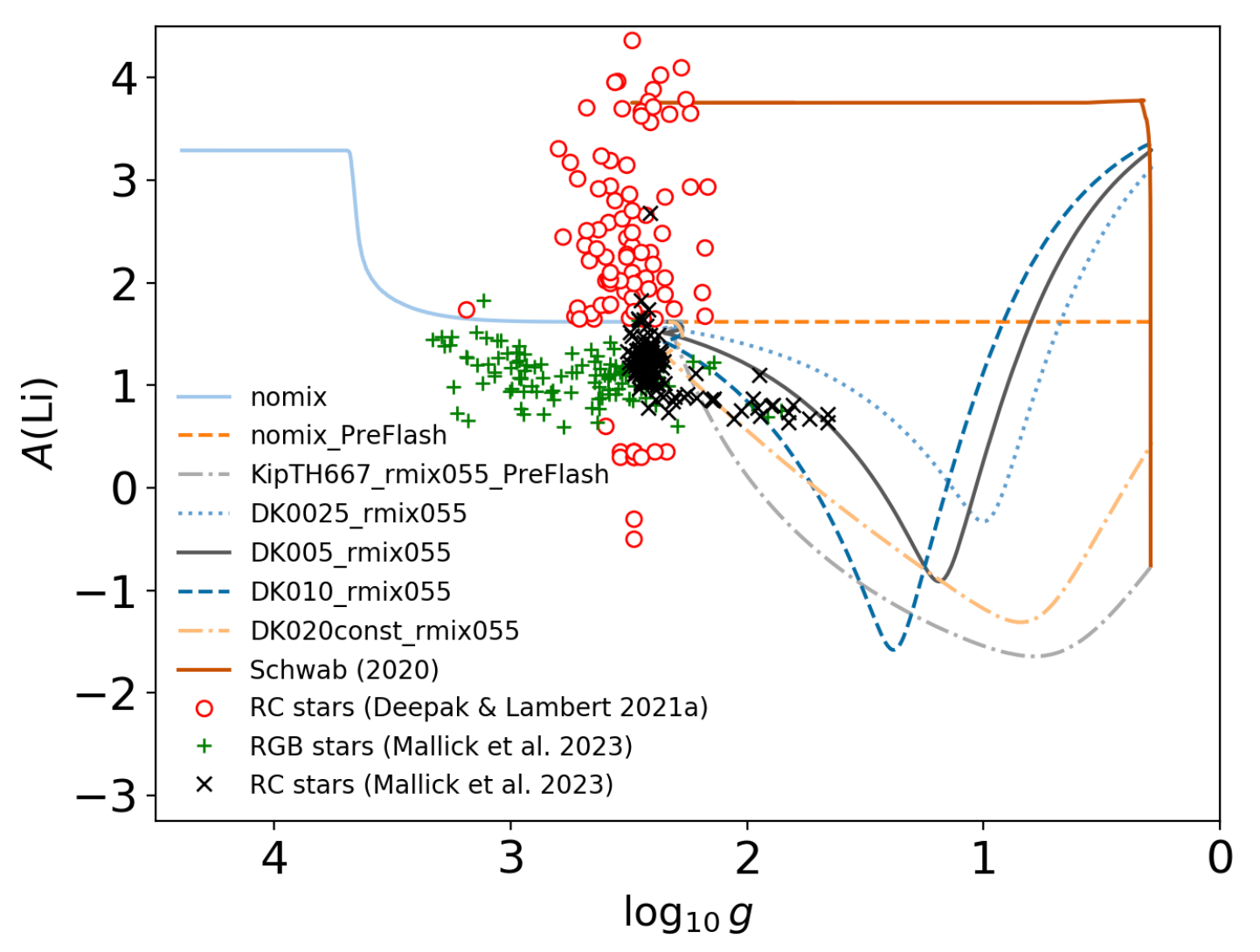}
  \caption{Similar to Figure \ref{fig:fig_li_rc}, but with the surface Li abundance plotted as a function of the gravity $g$.
           RGB and RC stars with $-0.5<\mathrm{[Fe/H]}<-0.1$ and $1 M_\odot < M < 1.4 M_\odot$ from \protect\cite{mallick2023}
           are added to this plot for comparison.
  }
  \label{fig:fig_li_lgg}
\end{figure}

\cite{schwab2020} put forward the hypothesis that the Li enrichment of RC stars could be a result of IGW mixing that
occurred in their precursors when they experienced the He-core flash at the RGB tip. 
Those IGWs could be excited by the first and strongest He-shell flash at its top convective boundary. However, for
the IGW mixing to be able to produce, via the Cameron-Fowler mechanism, sufficiently high Li abundances with $A(\mathrm{Li})\ga 1.5$,
as observed in many RC stars \citep[e.g.,][]{deepak2021}, during this relatively short event it has to be very fast.
We have repeated the calculations of \cite{schwab2020} for our $1.2\,M_\odot$ RGB-tip model using his
assumed values of IGW mixing rate and depth, namely, with the diffusion coefficient
$D_\mathrm{mix} = 10^{14}\ \mathrm{cm}^2\,\mathrm{s}^{-1}$ being kept constant in the radiative zone 
above the radius $r_\mathrm{mix}(X=0.5)$, where the H mass fraction has dropped to $X=0.5$, 
while the luminosity of the He-burning shell is exceeding $10^4\,L_\odot$. 
This extra-mixing setup and its corresponding evolution of the surface Li
abundance are shown in Figure~\ref{fig:fig_flash_rgb} and in Figures~\ref{fig:fig_li_rc} and \ref{fig:fig_li_lgg} (the solid brown line).

In order to understand if this hypothesis is plausible we
first note that its required IGW diffusion coefficient is significantly larger than the radiative diffusivity $K$ in the entire radiative zone
at $r > r_\mathrm{mix}(X=0.5)$ (compare the dashed blue and solid black curves in Figure ~\ref{fig:fig_flash_rgb}). 
This means that the IGW mixing should proceed on a dynamical,
rather than on a thermal timescale and, like convection, it should then dominate in the radial heat transport over the radiative diffusion, which
would change the thermal structure of the radiative zone, making it closer to adiabatic. As a result, the evolution of the star during
the He-core flash could change in a way that may have never been observed. Such drastic changes were discussed by \cite{denissenkov2012}
for low-mass RGB stars at the bump luminosity in which Li was assumed to be produced by extra mixing with $D_\mathrm{mix}\gg K$. 

Second, for IGW mixing to operate on a dynamical timescale, the waves should become non-linear and break. This could only happen
if the Richardson number associated with the IGW horizontal velocity shear $Ri\approx N^2/|\nabla\times\mathbf{u}|^2$ would drop below
its critical value $Ri_\mathrm{crit} =0.25$ \citep{press1981}. 
We can use the results of the 3D hydrodynamics simulation of IGW mixing in our RGB-tip model,
some of which are summarized at the end of Section \ref{sec:IGWs}, to demonstrate that this is unlikely to happen.
Indeed, from Figure 19 of \cite{blouin2023} we read a value of $D\sim 10^8\ \mathrm{cm}^2\,\mathrm{s}^{-1}$ for
the IGW diffusion coefficient estimated using the tracer fluid Gaussians at $r=450$ Mm. It is lower by a factor of 10 than a value of
the diffusion coefficient provided by Equation (\ref{eq:digw}) and the measured vorticities at the same radius. This difference vanishes if we reduce
the value of the parameter $\eta$ in Equation (\ref{eq:digw}) by the same factor, which is still a reasonable choice.
Now, if the scaling law $|\nabla\times\mathbf{u}|\propto L^{1/4}$ obtained for the bump luminosity model with IGWs
generated by the envelope convection can be applied to IGWs generated by the He-shell convection in the RGB-tip model then
from Equation (\ref{eq:digw}) for the ratio $\sim 10^6$ of the maximum luminosity of the first He-shell flash
to the luminosity of our star at the RGB tip we obtain an estimate of 
$D_\mathrm{IGW}\propto L^{1/2}\sim 10^3 D\sim 10^{11}\ \mathrm{cm}^2\,\mathrm{s}^{-1} \ll 
D_\mathrm{mix}\sim 10^{14}\ \mathrm{cm}^2\,\mathrm{s}^{-1}$. From the same equation
we can also estimate values of the IGW Richardson number $Ri\sim\eta (K/D_\mathrm{IGW})$.
For $K\approx 3.2\times 10^{12}\ \mathrm{cm}^2\,\mathrm{s}^{-1}$ at $r=450$ Mm taken from our RGB-tip model (Figure \ref{fig:fig_flash_rgb})
and $D_\mathrm{IGW}\sim 10^{14}\ \mathrm{cm}^2\,\mathrm{s}^{-1}$,
we obtain $Ri\approx 3.2$ and $Ri\approx 0.32$ for $\eta = 0.1$ and $\eta = 0.01$, respectively.
The second Richardson number is actually close to the critical value of $Ri_\mathrm{crit} = 0.25$, but it was obtained
for the He-shell luminosity at its maximum value of $\sim 10^9 L_\odot$, while the IGW mixing model
proposed by \cite{schwab2020} assumed that the value of $D_\mathrm{mix}\sim 10^{14}\ \mathrm{cm}^2\,\mathrm{s}^{-1}$ was maintained
all the time while the He-shell luminosity exceeded the much lower value of $10^4 L_\odot$, so the IGW mixing with this
high diffusion coefficient had to start operating with $Ri\approx 95\gg Ri_\mathrm{crit}$. We find this highly unlikely,
therefore we doubt that such fast IGW mixing is actually activated during the He-core flash. 

Furthermore, we have a doubt that the diffusion coefficient for IGW mixing can ever exceed the radiative diffusivity $K$, even if
breaking of IGWs produces turbulence. Indeed, because this happens in the radiative zone, we would expect that
the resulting turbulence should be strongly stratified at a low Prandtl number, in which case it is predicted that
$D_\mathrm{mix} < K$ \citep{lignieres2020,skoutnev2023}. 
Besides, with $D_\mathrm{IGW}\gg K$ the IGW mixing should transport heat
faster than radiation, thus transforming the radiative zone into an adiabatic one, which would make the propagation of
IGWs in it impossible. However, we have found that the value of $D_\mathrm{mix} = 0.99 K$ is not high enough to produce a significant amount of Li
in the same setup that \cite{schwab2020} used to model IGW mixing with $D_\mathrm{IGW}\gg K$ (the orange curve in Figure \ref{fig:fig_li_mth_radigw}).

\subsection{Lithium enrichment by rotation-induced mixing in binaries tidally locked on the RGB} 

\cite{denissenkov2006} proposed that tidal interaction of an RGB star with its close binary companion leading to a spin-up of that star via
synchronization of its rotational and orbital periods could enhance rotationally-induced mixing in its
radiative zone and, as a result, enrich its convective envelope in Li via the Cameron-Fowler mechanism.
\cite{casey2019} elaborated on that idea to argue that the Li-rich RC stars had been tidally spun-up on the RGB, then their internal rotation
was further accelerated during their contraction between the RGB tip and the RC phase with the total angular momentum conserved, and Li was produced by
enhanced rotational mixing. Their conclusion is based on the facts that most of the Li-rich giants are RC stars and that
the Li enrichment caused by enhanced extra mixing on the RGB cannot last longer than a few million years \citep{denissenkov2004}.
However, they seem to still admit the possibility that in the upper RGB stars not spun-up by tidal synchronization the mechanism for extra mixing is
thermohaline convection. Besides, their hypothesis does not explain why most, if not all, of the RC stars have higher Li abundances
than the ones predicted by the stellar evolution theory with RGB extra mixing. 

\section{Our hypothesis for the mechanism of RGB extra mixing and Li enrichment of RC stars}
\label{sec:our_hypo}

If \cite{casey2019} were right we would have to invoke at least three different angular-momentum and chemical-element transport mechanisms
in a same low-mass star, namely, something like the AMRI (Section \ref{sec:amri}), thermohaline and rotationally-induced mixing, 
to explain, respectively, their post-MS moderate core-envelope differential rotation, RGB extra mixing and its enhanced Li-producing mode. 
Here, applying the principle of Occam's razor, we discuss a possibility 
that a same physical mechanism is responsible for both the coupled evolutionary changes of
the core and envelope rotation in low-mass subgiant and early-RGB stars, as measured by asteroseismology, and the evolutionary declines of
the $^{12}$C/$^{13}$C and C/N ratios in upper RGB stars, as revealed by stellar spectroscopy. Furthermore, we speculate that
the same RGB extra mixing mechanism that begins to manifest itself at the bump luminosity and initially leads to a further depletion of the surface Li
abundance in upper RGB stars gets enhanced and begins to produce Li in all these stars when they approach the RGB tip.

Initially, we focused only on the upper RGB extra mixing and considered the following two candidates for its mechanism: 
mixing by small-scale turbulence driven by horizontal velocity shear produced by internal gravity waves (IGWs), as described
in Section \ref{sec:IGWs}, and the joint operation of rotation-driven meridional circulation and turbulent diffusion, as
described by DT00.
We have excluded thermohaline convection from our consideration because it results in a strong depletion of the surface Li abundance in upper RGB stars
(the dot-dashed gray curve in Figure \ref{fig:fig_li_rc}),
even when its diffusion coefficient is artificially magnified by a large factor of $\sim 200$ not supported by hydrodynamics simulations
\citep{denissenkov2010,traxler2011,denissenkov2011}, and
because it is not clear how it can be enhanced and produce Li near the RGB tip to support the main idea of this work.
Diffusion coefficients corresponding to these mechanisms, including thermohaline convection,
can all be represented as $D_\mathrm{mix} = fK$, where $f=f(L_\mathrm{bce})$ for mixing by IGWs generated by the envelope convection,
$f=f(\Omega_\mathrm{bce},(\partial\Omega/\partial r)_\mathrm{rad})$ for rotation-driven mixing,
and $f=f(X_\mathrm{bce}(^3\mathrm{He}),(\partial X(^3\mathrm{He})/\partial r)_\mathrm{rad})$ for thermohaline convection,
where the parameters $(\ldots)_\mathrm{bce}$ and $(\ldots)_\mathrm{rad}$ refer to the base of the convective envelope and
to the radiative zone, and $L_\mathrm{bce} = L$. 

Here, we use the same MESA model of a low-mass star with the initial mass $1.2M_\odot$ and metallicity [Fe/H]\,$=-0.3$ that
was introduced in Section \ref{sec:thm} as a representative for the Li-rich RC stars and whose evolution from the zero-age MS through 
to the thermally-pulsing AGB phase was shown in Figure \ref{fig:fig1}. Like for the case of thermohaline convection, the mixing depth
is fixed at the radius $r_\mathrm{mix} = 0.055R_\odot$ in our parametric models of RGB extra mixing. The initial value of
the factor $f$ at the bump luminosity $f(L_\mathrm{bump})$ is treated as a free parameter. Our goal is to find reasonable physically-motivated scalings of
$f$ with the luminosity, e.g. a power-law form of $f(L) = f(L_\mathrm{bump})(L/L_\mathrm{bump})^p$ with $p>1$,
for IGW, rotationally-induced, and AMRI (Section \ref{sec:amri}) mixing mechanisms and demonstrate that with these scalings they can produce 
the surface Li abundances in our models, by the time they will reach the RGB tip, comparable to those measured in RC stars.
We artificially limit the value of $f\le 1$, so that extra mixing does not change thermal stratification of the radiative zone.

Obviously, thermohaline mixing does not satisfy the requirement of our hypothesis that its corresponding factor $f$
should increase with the luminosity on the upper RGB, so that Li is produced by the Cameron-Fowler mechanism 
when the factor $f$, and therefore the RGB extra mixing diffusion coefficient $D_\mathrm{mix} = fK$, increases in
the star approaching the RGB tip. Instead, $f=f(X_\mathrm{bce}(^3\mathrm{He}),(\partial X(^3\mathrm{He})/\partial r)_\mathrm{rad})$
decreases when a low-mass star climbs the upper RGB following a decline of both $X_\mathrm{bce}(^3\mathrm{He})$ 
and $(\partial X(^3\mathrm{He})/\partial r)_\mathrm{rad}$ caused by the $^3$He burning in the vicinity of the H-burning shell
and RGB extra mixing. In Figure \ref{fig:fig_li_rc}, its predicted Li-destruction/production curve is compared
with the one for the model with $D_\mathrm{mix} = 0.02K$ and the same value of $r_\mathrm{mix} = 0.055R_\odot$.
Both models actually begin to produce some Li near the RGB tip simply because
the radiative diffusivity $K$ is proportional to $L$, but the latter model ({\tt DK02const\_rmix055}) makes more Li because it has a constant value of $f=0.02$,
whereas $f$ decreases with $L$ in the former model.

\subsection{IGW mixing on the upper RGB enhanced by the increasing luminosity}
\label{sec:igw_enhanced_mix}

At the beginning of this work, before we had done the corresponding hydrodynamics simulations, we thought that
IGWs could provide a mechanism for both the RGB extra mixing and Li enrichment by its enhanced efficiency at the RGB tip. Our line of reasoning  
was based on the following data. Under certain assumptions, it can be anticipated that the horizontal and vertical components of
IGW velocity, and therefore its vorticity in Equation (\ref{eq:digw}), are all proportional to $L^{2/3}$ \citep[e.g., Section 5.2 in][]{herwig2023}.
Hence, the diffusion coefficient for IGW mixing driven by the IGW horizontal velocity shear was expected to increase
with the luminosity as $D_\mathrm{IGW} = f(L) K$, where $f(L)\propto L^{4/3}$.
\cite{schwab2020} needed a diffusion coefficient $D_\mathrm{mix}\sim 10^{14}\ \mathrm{cm}^2\mathrm{s}^{-1}$ for
IGW mixing triggered by the first He-shell flash at the RGB tip and then maintained at this high value for luminosities between
$L_\mathrm{He}\sim 10^4 L_\odot$ and $L_\mathrm{He}\sim 10^9 L_\odot$ to be able to produce Li in amounts comparable to
those observed in RC stars. If we take a middle value of the He-shell luminosity 
$L_\mathrm{He,mid}\sim 10^6 L_\odot$ and use the power law $D_\mathrm{IGW}\propto L^{4/3}K$ then we obtain estimates of 
$D_\mathrm{IGW}(L_\mathrm{tip})\sim 10^{14} (L_\mathrm{tip}/L_\mathrm{He,mid})^{4/3}\approx 2.5\times 10^{10}\ \mathrm{cm}^2\mathrm{s}^{-1}$ 
and $D_\mathrm{IGW}(L_\mathrm{bump})\sim 10^{14} (L_\mathrm{bump}/L_\mathrm{He,mid})^{4/3}\approx 2.5\times 10^8\ \mathrm{cm}^2\mathrm{s}^{-1}$
at the RGB tip and bump luminosities, $\log_{10}(L_\mathrm{tip}/L_\odot) = 3.3$ and $\log_{10}(L_\mathrm{bump}/L_\odot) = 1.8$, respectively. 
These IGW diffusion coefficients would be sufficiently large to explain the RGB extra mixing \citep{denissenkov2003}.
Moreover, extra mixing on the upper RGB with the diffusion coefficient $D_\mathrm{mix} = D_\mathrm{IGW} = f(L)K$ could reproduce
the evolutionary declines of [C/N] and $^{12}$C/$^{13}$C as well as the Li enrichment of RC stars for the fixed mixing depth 
$r_\mathrm{mix} = 0.055R_\odot$ and $f(L) = f(L_\mathrm{bump})(L/L_\mathrm{bump})^{4/3}$ with $f(L_\mathrm{bump}) = 0.0025$,
$f(L_\mathrm{bump}) = 0.005$, and $f(L_\mathrm{bump}) = 0.010$ (the dotted blue, solid black, and dashed blue curves 
in Figures \ref{fig:fig_cn}, \ref{fig:fig_c12c13}, and \ref{fig:fig_li_rc}).
Note that on the Li abundance plot these three models arrive at values of $A(\mathrm{Li})\ga 0.5$ only when $\log_{10}(L/L_\odot)\ga 2.75$
where there are almost no measurements of the Li abundance in field low-mass stars \citep[e.g., Figure 7 in][]{deepak2021}
to verify or reject our hypothesis. 

However, our recent 3D hydrodynamics simulations of convection and IGWs in the $1.2 M_\odot$ upper RGB star
\citep{blouin2023} have shown that even in its highest-luminosity RGB-tip model the efficiency of IGW mixing is much lower than the
observationally constrained rates of the RGB extra mixing and that
the vorticity of IGW motion in the radiative zone is proportional to $L^{1/4}$ (again, if this scaling law obtained for
the bump luminosity model can be applied to the RGB-tip model), 
rather than to $L^{2/3}$ as we anticipated.
Therefore, IGW mixing is probably the wrong mechanism for our Li-enrichment hypothesis to work.

\subsection{Rotationally-induced mixing on the upper RGB enhanced by the increasing mass loss}
\label{sec:sec_ml_enhanced_mix}

When a rotating star does not lose any mass, and therefore angular momentum, the competition between angular momentum transport
by the rotationally-induced meridional circulation and turbulent diffusion in its radiative zone, as described by Equation (\ref{eq:amt}),
may reach a state of equilibrium \citep[e.g.][]{denissenkov1999,denissenkov2000b}. 
\cite{zahn1992} suggested that such equilibrium could be broken by a magnetized stellar wind and its associated
strong angular momentum loss by the star, in which case the angular velocity profile $\Omega(r)$ for the shellular
rotation in the radiative zone would become steeper and that would enhance the efficiency of rotational mixing.
For the case of moderate wind, \cite{zahn1992} obtained the following estimate of an effective diffusivity for
rotational mixing in the asymptotic regime:
\begin{eqnarray}
D_\mathrm{eff} = \frac{C_\mathrm{h}}{50}\frac{r|U|}{\alpha} = \frac{C_\mathrm{h}}{20}\frac{\Omega_\mathrm{s}}{\Omega(r)}
\frac{k^2}{\alpha}\frac{R^2}{t_\mathrm{J}}\frac{\rho_\mathrm{m}}{\rho},
\label{eq:deff}
\end{eqnarray}
where $C_\mathrm{h}\la 1$ is a free parameter, $\alpha = d\ln (r^2\Omega)/d\ln r$, $\Omega_\mathrm{s} = \Omega(R)$ the surface angular
velocity, $t_\mathrm{J} = k^2 MR^2\Omega_\mathrm{s}/(-dJ/dt)$ the timescale of the angular-momentum loss with
$k^2 = (2/3)\int r^2 dM_r/(MR^2)$ representing a dimensionless moment of inertia of the star, 
and $\rho_\mathrm{m} = 3M_r/(4\pi r^3)$ the mean density of a sphere of the radius $r$.
\cite{charbonnel1995} used Equation (\ref{eq:deff}) in a model of RGB extra mixing that was able
to successfully reproduce the observed $^{12}$C/$^{13}$C ratios in globular-cluster and field population-II
upper RGB stars simultaneously with an initial strong depletion of their surface Li abundances.

Our hypothesis assumes that Li enrichment of RC stars occurs in their progenitors when they approach the RGB tip
where large amounts of Li are produced via the Cameron-Fowler mechanism by enhanced RGB extra mixing.
For IGW mixing such enhancement could be a direct consequence of the increasing luminosity
but, according to the results of 3D hydrodynamics simulations of \cite{blouin2023}, this does not seem
to be the case (Section \ref{sec:igw_enhanced_mix}). Alternatively, if it could still be
possible to associate RGB extra mixing with rotationally-induced meridional circulation and turbulent diffusion, as we
discussed in Section \ref{sec:rotmix}, then Equation (\ref{eq:deff}) would provide a mechanism for
the RGB extra mixing to get enhanced near the RGB tip. Indeed, this equation can be written in the following form: 
\begin{eqnarray}
D_\mathrm{eff} = \frac{C_\mathrm{h}}{20\alpha}\frac{\Omega_\mathrm{s}}{\Omega(r)}
\frac{R^2}{M}\frac{\rho_\mathrm{m}}{\rho}\left(-\frac{dM}{dt}\right),
\label{eq:deff_dmdt}
\end{eqnarray}
where $dM = dJ/(R^2\Omega_\mathrm{s})$ is a mass lost with the angular momentum $dJ$.
There is a number of different prescriptions for the mass-loss rate $dM/dt$ by low-mass stars on the RGB.
Six of them are listed in Table 1 of \cite{catelan2009} and presented in his Figure 4
in the form of an integrated mass loss along the RGB as a function of metallicity [Fe/H] for
a fixed age of 12 Gyr \citep[also, see Appendix in][]{catelan2000}.
The widely-used Reimers formula $dM/dt = -4\times 10^{-13}\eta_\mathrm{R}\,(L/gR)\ M_\odot\,\mathrm{yr}^{-1}$ \citep{reimers1975},
where the surface luminosity $L$, radius $R$, and gravity $g$ are all expressed in solar units, and $\eta_\mathrm{R}$ is a free parameter
(we use the value of $\eta_\mathrm{R} = 0.36$),
gives the minimum RGB-integrated mass loss. For this prescription, our stellar evolution calculations predict the scaling relation
$dM/dt\propto L^{5/3}\propto L^{1/3}K$, since $K\propto L$, and therefore $D_\mathrm{eff}\propto L^{1/3}K$.
However, if we take the modified Reimers mass-loss rate from Table 1 of \cite{catelan2009},
$dM/dt\propto (L/gR)^{+1.4}$, that gives the second-largest RGB-integrated mass loss, 
then from the result obtained for the standard Reimers rate we immediately find that 
it increases with the luminosity as $dM/dt\propto L^{7/3}\propto L^{4/3}K$,
and $D_\mathrm{eff}\propto L^{4/3}K$. This is similar to our anticipated scaling relation for
the IGW diffusion coefficient that we discussed in the previous section, therefore the results of our
stellar evolution computations with the RGB extra mixing diffusion coefficient
$D_\mathrm{mix} = f(L_\mathrm{bump})(L/L_\mathrm{bump})^{4/3}K$ that are presented there, in particular the dotted blue, solid black, and dashed blue
Li-destruction/production curves in Figure \ref{fig:fig_li_rc}, can directly be applied here. 

Note that in the case of a magnetized stellar wind $dJ$ can significantly exceed $(R^2\Omega_\mathrm{s})dM$
because its magnetic coupling with the stellar envelope can extend to distances $r\gg R$, then
even the standard Reimers mass-loss rate may be accompanied by a sufficiently strong loss of angular momentum
to make Zahn's rotationally-induced mixing in the asymptotic regime of moderate stellar wind
fast enough to produce large amounts of Li at the RGB tip.

\subsection{AMRI as a possible mechanism for the RGB extra mixing and Li enrichment of RC stars}
\label{sec:amri_model}

In Section \ref{sec:amri}, we suggested that the azimuthal magneto-rotational instability (AMRI) could ae the right
physical mechanism not only for the angular-momentum transport in low-mass stars but also for mixing in their radiative zones
on the upper RGB. The dependence of its diffusion coefficient (\ref{eq:DAMT}) on the ratio of the mean angular velocities of
the radiative core $\Omega_\mathrm{core}$ and convective envelope $\Omega_\mathrm{env}$ 
indicates that the AMRI mixing could be significantly enhanced towards the RGB tip.
Indeed, between the bump luminosity and the RGB tip the radius of our $1.2 M_\odot$ model star increases by an order of magnitude
on a timescale of 45 Myr. If the characteristic timescale of core-envelope rotational coupling \citep[e.g.,][]{denissenkov2010b} 
is longer than this then rotation of the expanding convective envelope would slow down by a factor of $100$, because of
the conservation of its angular momentum, and
the AMRI diffusion coefficient might increase by a factor of $10^4$. That would be sufficient for the AMRI mixing
to produce large amounts of Li at the RGB tip, provided that it could start to manifest itselfs as the RGB extra mixing at the bump luminosity.
Obviously, the efficiency of the AMRI mixing on the upper RGB should depend not only on the $\Omega_\mathrm{core}/\Omega_\mathrm{env}$ ratio, but also
on the initial rotational velocity of the star as well as on its mass and metallicity.

\begin{figure}
  \centering
  \includegraphics[width=\columnwidth]{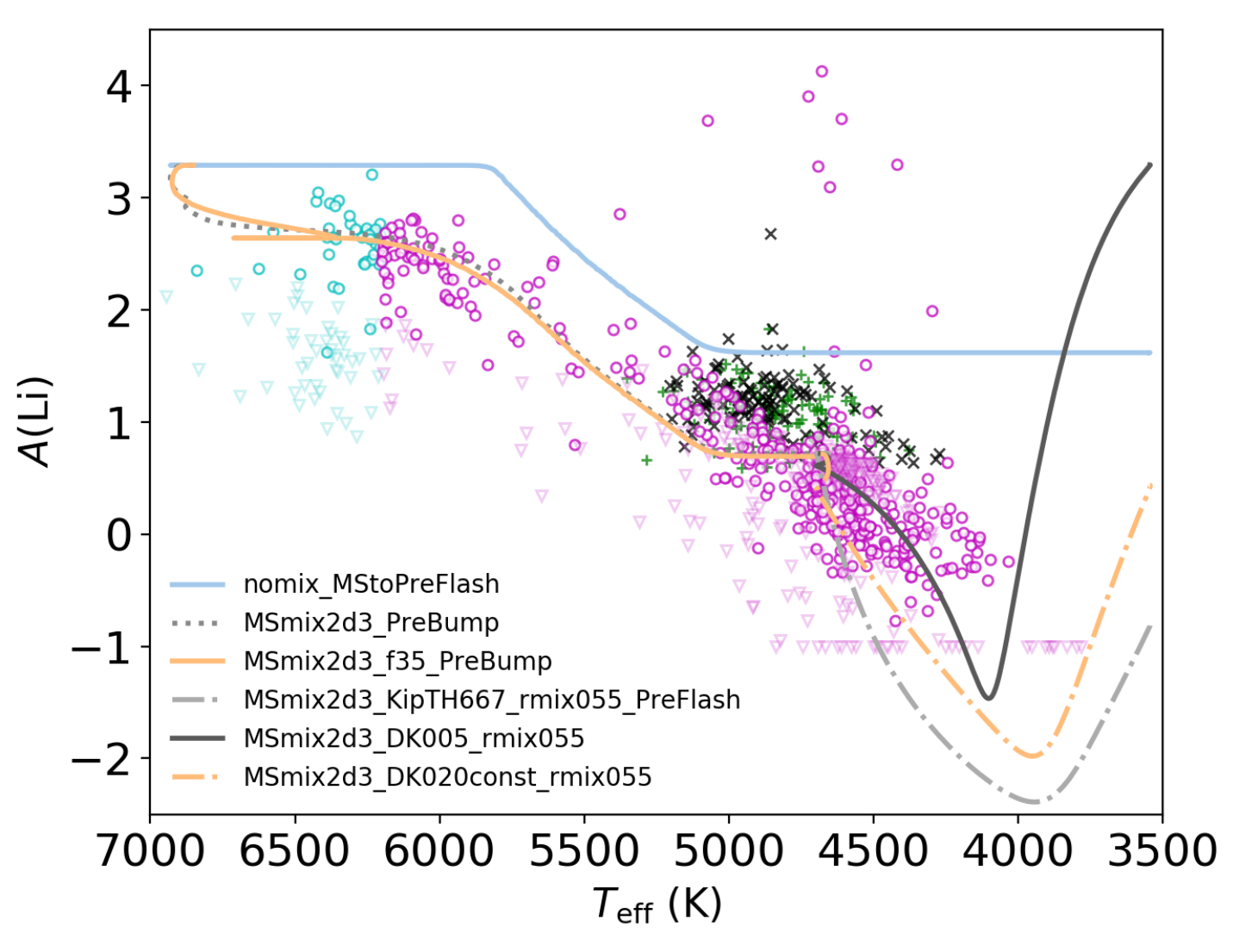}
  \caption{Circles are Li abundances and triangles are their upper limits from \protect\cite{magrini2021} 
           for MS turn-off (cyan), SGB, and RGB (magenta) stars with $-0.5<\mathrm{[Fe/H]}<-0.1$ and 
           $1 M_\odot < M < 1.4 M_\odot$, and crosses are the same RGB and RC stars from \protect\cite{mallick2023} that are displayed in Figure
           \ref{fig:fig_li_lgg}. The dotted black curve shows Li abundances predicted by our $1.2 M_\odot$ model with $\mathrm{[Fe/H]}=-0.3$,
           in which we have included extra mixing on the MS with $D_\mathrm{mix} = 2\times 10^3\ \mathrm{cm}^2\mathrm{s}^{-1}$ operating only down
           to the depth where $T=3.75\times 10^6$ K to reproduce the observed depletion of the surface Li and Be abundances in MS
           low-mass stars, as explained in Section \ref{sec:sec_amt_li_ms}. The solid orange curve represents a model in which we have added
           mixing beyond the boundary of the H convective core (overshooting) decaying exponentially with
           the distance from the core on the length scale $f_\mathrm{ov} = 0.035$ pressure scale height. The blue curve is the model without
           any extra mixing. The last three models are counterparts of their corresponding models from Figure \ref{fig:fig_li_lgg}, except
           that these include MS extra mixing. 
  }
  \label{fig:fig_li_teff}
\end{figure}

\begin{figure}
  \centering
  \includegraphics[width=\columnwidth]{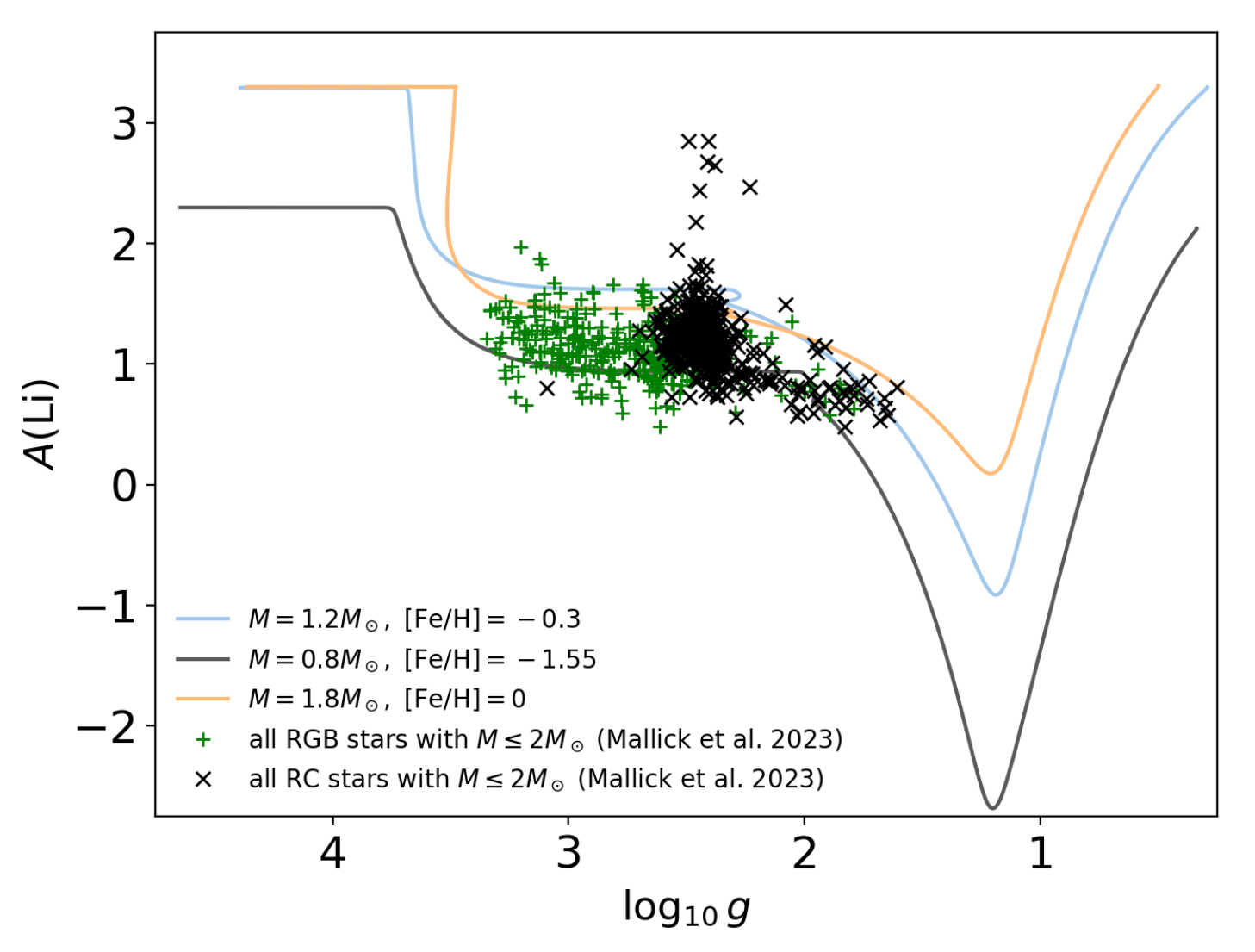}
  \caption{Changes of the surface Li abundance caused by the FDU and enhanced extra mixing on the upper RGB modelled using the diffusion coefficeint
           $D_\mathrm{mix} = 0.005 (L/L_\mathrm{bump})^{4/3} K$ in stars with different initial masses and metallicities.
  }
  \label{fig:fig_li_diff_m}
\end{figure}

\subsection{Angular momentum transport and depletion of the surface Li abundance in MS low-mass stars}
\label{sec:sec_amt_li_ms}

The problem of angular momentum transport (AMT) in low-mass stars after they have left the MS and the following RGB extra mixing 
resembles a similar problem of AMT and
evolutionary decline of the surface Li abundance in their MS precursors. We have already mentioned in Introduction that the Li abundance
in the atmospheres of MS solar-type stars is observed to be decreasing with their age by more than two orders of magnitude
\citep{sestito2005,carlos2019,rathsam2023}. Besides having its surface Li abundance reduced to a value of $A(\mathrm{Li})=1.1$ from
the initial abundance of $A(\mathrm{Li})=3.3$ \citep{asplund2009}, the Sun has a radiative core rotating uniformly,
at least down to the radius of $\sim 0.2R_\odot$, as revealed by helioseismology \citep{couvidat2003}. 
\cite{dumont2021} have shown that rotational mixing and AMT
cannot reproduce the observed Li depletion in the Sun simultaneously with the solid-body rotation of the solar core, without
adding a constant viscosity $\nu_\mathrm{add} = 2.5\times 10^4\ \mathrm{cm}^2\mathrm{s}^{-1}$ and a diffusion coefficient whose value 
at the top of the radiative core has increased from $D_\mathrm{mix}\sim 10^2\ \mathrm{cm}^2\mathrm{s}^{-1}$ to 
$D_\mathrm{mix}\sim 10^4\ \mathrm{cm}^2\mathrm{s}^{-1}$ over the age of the Sun
and is exponentially decreasing with the depth, as prescribed by equation (2) from \cite{richard2005}. These additional transport
coefficients are used to model turbulence of unknown origin with their amplitudes considered as free parameters adjusted to
reproduce observations. More recently, \cite{dumont2023} has demonstrated that the same model that \cite{dumont2021} used to simulate 
the AMT and Li depletion in MS solar-type stars can also reproduce observational changes of the core and envelope averaged
angular velocities between the subgiant and red giant evolutionary phases 
in 8 stars with metallicities $-0.4<\mathrm{[Fe/H]}<-0.4$ and masses $1.1 M_\odot < M < 1.5 M_\odot$,
provided that the additional viscosity depends on the differential rotation as prescribed by equation (\ref{eq:DAMT})
with the parameters $D_0 = 7.5\times 10^2\ \mathrm{cm}^2\mathrm{s}^{-1}$ and $\alpha = 1.5$. \cite{dumont2023} has also
confirmed the conclusion made earlier by \cite{deheuvels2020} that relatively low values of the ratio
$\Omega_\mathrm{core}/\Omega_\mathrm{env}$ measured in low-mass stars at the beginning of the subgiant branch (SGB) indicate
that there is a sharp decrease of the AMT efficiency, nearly by one order of magnitude, at this evolutionary phase,
after which AMT is well described by the AMRI-like viscosity law.

Therefore, an additional AMT driving mechanism, different from pure rotational and AMRI, is required to explain rotation of low-mass stars
on the MS, which should become less efficient than AMRI on the SGB. Such mechanisms could be the Taylor-Spruit dynamo \citep{spruit2002}
or IGWs \citep{ringot1998}. Indeed, \cite{eggenberger2022} and \cite{charbonnel2005} showed that the AMT and mixing driven, respectively,  
by the Taylor-Spruit dynamo and by IGWs could reproduce both rotation and surface Li abundance in the Sun, while \cite{cantiello2014} and \cite{fuller2014} came
to the conclusions that neither the Tayler-Spruit dynamo nor IGWs could explain the spin-down of core rotation in low-mass stars on the RGB.
However, \cite{denissenkov2008b} demonstrated that asymmetric attenuation of prograde and retrograde IGWs should produce large-scale
oscillations in the rotation profile of the solar core that are not observed, which would make IGWs an unlikely mechanism of AMT
in solar-type stars. 

To see if the depletion of Li in low-mass stars on the MS can affect our hypothesis of their enrichment in Li by enhanced extra mixing on the upper RGB
we have simulated the first of these processes in our $1.2 M_\odot$ model star using a diffusion coefficient $D_\mathrm{mix} = 2\times 10^3\ \mathrm{cm}^2\mathrm{s}^{-1}$,
which turns out to accidentally have a value intermediate between the minimum and maximum amplitudes of 
the additional mixing coefficient that \cite{dumont2021} adjusted to reproduce
the decline of the Li abundance in MS solar-type stars, and we have assumed that it remains constant down to a depth where the temperature reaches
$T_\mathrm{mix} = 3.75\times 10^6$ K, below which it vanishes. With these values of $D_\mathrm{mix}$ and $T_\mathrm{mix}$ we were able to reproduce the 
Li abundances reported by \cite{magrini2021} for Milky-Way field MS turn-off, SGB, and RGB stars with metallicities and masses close to those used for our model
(dotted black curve in Figure~\ref{fig:fig_li_teff}). 
The value of the parameter $T_\mathrm{mix}$ was selected to also result in an evolutionary depletion of the surface Be abundance 
in our model resembling those presented in Figure 9 of \cite{dumont2023} for his $1.2 M_\odot$ models.
The solid orange curve in Figure~\ref{fig:fig_li_teff} displays the evolution of
the surface Li abundance in a model that also includes boundary mixing (overshooting) at the H convective core with a diffusion coefficient exponentially decreasing 
with a distance from the convective boundary on the length scale of
$f_\mathrm{ov} = 0.035$ pressure-scale height, as in \cite{bocek2019}. The purpose of this model was to simulate the extension of
the MS caused by rotation that is not included in our computations. Comparison of the dotted black and solid orange curves with the solid blue
curve representing a standard model without extra mixing and comparison of the first two curves with the observational data shows
that by the time the low-mass stars arrive at the bump luminosity ($T_\mathrm{eff}\approx 4700$ K) their surface Li abundance is decreased by
one to two orders of magnitude compared to what would be expected for the standard evolution. However, this does not affect the following
enhancement of the surface Li abundance in our model, as illustrated by the dot-dashed gray, solid black, and dot-dashed orange curves
in Figure~\ref{fig:fig_li_teff} that represent counterparts of their corresponding models from Figures~\ref{fig:fig_li_rc} and \ref{fig:fig_li_lgg}.

\section{Discussion}
\label{sec:concl}

The problem of AMT and extra mixing in low-mass stars should ideally be discussed in the context of an accurate numerical solution of 
a system of non-linear magnetohydrodynamics (MHD) equations for a three-dimensional $4\pi$ geometry with a structure of a low-mass star at
its different evolutionary phases between the zero-age MS and HB used as a background for its initial setup. Unfortunately, such simulations are not possible
to do at present with a sufficiently high resolution and for a sufficiently long period of the star lifetime. Therefore, they are usually
replaced by simulations in which complexity of the problem is reduced by going from three to two dimensions, changing the spherical geometry
to a box- or wedge-shaped one, and ignoring some ingredients of this multi-physics problem, e.g. magnetic fields, rotation, sound waves,
radiative diffusion, etc. We do believe that even such simplified simulations still provide useful information about AMT and extra mixing
in the star and their results should be taken into account when comparing model predictions with observational data. However, it is not clear
to what degree we can trust their results and conclusions. In this situation, we consider it is acceptable to employ
simple parametric semi-empirical models to see which better reproduces observational data and for which values of its parameters,
and then choose a reduced MHD model with transport properties closest to those suggested by the parametric model. An example of this approach that we
have utilized in this work is the parametric prescription (\ref{eq:Dth}) for the thermohaline diffusion coefficient derived
from linearized Boussinesq equations \citep[e.g.,][]{denissenkov2010}. In this parametric model, the mixing rate is proportinal to
the square of a salt-finger length to diameter ratio $a$. \cite{charbonnel2007} have shown that the evolutionary changes of
the Li, C, and N elemental abundances and $^{12}\mathrm{C}/^{13}\mathrm{C}$ isotopic ratio in upper RGB stars can be reproduced by this model assuming
that $a\approx 7$. That prediction motivated \cite{denissenkov2010}, \cite{denissenkov2011}, and \cite{traxler2011} to perform 2D and 3D numerical simulations of
thermohaline convection in radiative zones of upper RGB stars. They all found a value of $a\approx 1$, which cast doubt on thermohaline
convection as the mechanism of RGB extra mixing. The fact that RGB extra mixing starts near the bump luminosity, where the local $\mu$ inversion
develops in the vicinity of the H-burning shell, does not prove yet that thermohaline convection driven by the negative $\mu$ gradient
is the right mechanism of RGB extra mixing because the $\nabla\mu\leq 0$ profile in the radiative zone can facilitate operation of any extra mixing
in it by reducing the stabilizing effect of the  buoyancy force. Now, \cite{fraser2024} have shown that a radial magnetic field
of the magnitude $100$ G to $1000$ G can stabilize the growth of salt fingers in the radiative zone, in which case a value of $a\approx 7$
can be achieved there, and nearly at the same time asteroseismology has detected radial magnetic fields of the order of $100$ kG in 26 {\it Kepler} low-mass red giants.
Therefore, we have made a conclusion that such magnetically-assisted thermohaline convection deserves to be considered as a possible mechanism of RGB extra mixing
alternative to the AMRI.

In a real star, rotation, magnetic fields, growth of salt fingers at $\nabla\mu < 0$, IGWs, convective boundary mixing (overshooting)
are expected to interact with each other, which may affect AMT and extra mixing produced by various MHD processes individually. But,
in order to accurately estimate effects of these interactions and their magnitudes we would again need to find a solution of
the system of MHD equations with all these ingredients plus appropriate boundary conditions, e.g. angular momentum loss via magnetized stellar wind, 
included, which is not possible yet. In the absence of such solution
we can rely only on conclusions made using simplified hydrodynamics and MHD models. 

For example, \cite{varghese2024} have studied
the impact of rotation on mixing by IGWs generated by the H-core convection in a $7 M_\odot$ MS star of solar metallicity.
They have solved the Navier-Stokes equations in the anelastic approximation in a 2D equatorial slice of the star similar
to how it was done by \cite{rogers2013}. After that, they have estimated the diffusion coefficient for IGW mixing with the method of tracer
particles that was earlier developed and used by \cite{rogers2017}. The main conclusion made by \cite{varghese2024} is that the rate of 
IGW mixing decreases with increase in rotation. They attribute this effect to the impact rotation makes on convection that excites
IGWs at its boundary with the radiative zone, which reduces amplitudes of the excited waves. This result indirectly confirms the finding of
more strong damping of gravito-inertial waves near their excitation region by \cite{andre2019}.
However, we do not know how credible the conclusion of \cite{varghese2024} is, given that 
they used a model with a radiative diffusivity $K$ increased by four orders of magnitude compared to its actual value with
a correspondingly increased heating in the H convective core, the Prandtl number $\mathrm{Pr} = \nu/K = 80$ by many orders of
magnitude exceeding its values in stellar radiative zones, and they did not discuss the physics behind their observed IGW mixing
and scaling of ther estimated IGW diffusion coefficient to realistic values of stellar parameters. It is possible that
the surprisingly high efficiency of IGW mixing reported by them was caused by the assumed extremely high viscosity $\nu = 4\times 10^{13}\ \mathrm{cm}^2\mathrm{s}^{-1}$.

Here is another example.
One of the most elaborated methods for computations of 1D stellar evolution with rotationally-induced AMT and mixing is probably
the one used by \cite{dumont2021}. Still, they need to assume that in low-mass MS stars there are additional processes 
significantly contributing to AMT and chemical mixing and, because their mechanisms are not known, \cite{dumont2021} do not discuss
how they may affect the operation of other transport processes included in their model.

In this work we have discussed, in a more or less detailed way, several mechanisms of extra mixing in the radiative zones of low-mass stars on the upper RGB,
above the bump luminosity, including the rotationally-induced meridional circulation with turbulent diffusion and their enhanced
mixing mode in the asymptotic regime of moderate stellar wind, the AMRI, thermohaline convection
and its magnetically-assisted mode, buoyancy of magnetic flux tubes, the Tayler-Spruit dynamo,
and mixing by internal gravity waves. We have suggested that the AMRI may potentially
explain not only the angular momentum transport (AMT) between the rapidly-rotating radiative cores and the slower-rotating convective 
envelopes in these stars during their evolution from the MS turn-off through to the HB, as was previously revealed by asteroseismology 
\citep[e.g.,][]{dumont2023,spada2016,moyano2023}, but also the RGB extra mixing and probably even its enhanced mode near the RGB tip
that may be responsible for the Li-enrichment of RC stars. 

It is interesting that approximately at the same time, in the autumn of 2021, when we had started to develop our idea of Li production
by the Cameron-Fowler mechanism with IGW mixing in low-mass stars on the upper RGB, assuming that its rate should increase with the luminosity as
$D_\mathrm{mix} = f(L_\mathrm{bump})(L/L_\mathrm{bump})^{4/3}K$, \cite{li2023} had submitted a paper where they put forward a similar hypothesis. However,
instead of mixing by IGWs caused by their produced horizontal velocity shear that has led us to the above scaling relation, they used
the model of IGW mixing based on the consideration of a random walk of tracer particles pushed by an ensemble of IGWs and 
assisted by radiative heat diffusion that had been proposed and developed by \cite{montalban1994} and \cite{montalban2000}. Like in our case,
it is not a surprise that for their IGW mixing they have found a diffusion coefficient increasing with the luminosity along the upper RGB, since
the radiative diffusivity and kinetic energy of IGWs are both increasing with $L$. Because, following \cite{kumar2020}, they
adopted a very low efficiency of thermohaline convection with $\alpha_\mathrm{th} = 100$, which corresponds to the finger aspect ratio
$a\approx 1$ supported by the hydrodynamics simulations of \cite{denissenkov2010}, \cite{traxler2011}, and \cite{denissenkov2011}, they had to adjust
parameters for their model of IGW mixing that would make it as efficient as the RGB extra mixing. Therefore, in this
respect their model is also similar to our parametric model of IGW mixing for which we have used the observationally constrained values of
the parameter $f(L_\mathrm{bump})$. However, we had decided to postpone the completion of our work until the results of our 3D hydrodynamics
simulations of convection and IGWs in our $1.2 M_\odot$ upper RGB model star would be obtained and published \citep{blouin2023}. These results have shown that
even in the highest-luminosity RGB-tip model the rate of IGW mixing is much slower than what is required to identify it
with the RGB extra mixing, therefore we have rejected IGWs as the main mechanism of extra mixing in upper RGB stars and
Li-enrichment of RC stars.

Our $1.2 M_\odot$ model star spends only $14\%$ of its upper RGB lifetime (6 Myr out of 44 Myr) between the luminosities 
$\log_{10}\,(L/L_\odot)\approx 2.75$ and $\log_{10}\,(L/L_\odot) = 3.3$, where $A(\mathrm{Li})$ in its atmosphere
becomes positive and continues to grow in our parametric model of RGB extra mixing (Figure \ref{fig:fig_li_rc}).
Values of $A(\mathrm{Li})$ exceed $1.5$, i.e. the star becomes a Li-rich red giant, when compared with its post-FDU Li abundance for the evolution
without extra mixing on the MS,
only at $\log_{10}\,(L/L_\odot)\ga 2.9$ during the last $9\%$
of its upper RGB lifetime (4 Myr out of 44 Myr). These lifetimes are reduced to $6\%$ and $4\%$, respectively, if we compare
them with the RGB evolutionary time between the end of the FDU and the RGB tip (100 Myr). The time our star has $A(\mathrm{Li})>1.5$ on the upper RGB is only $7\%$
of its HB lifetime (56 Myr), which is not very different from $17\%$ of Li-rich objects among the ``highly-evolved AGB or RGB stars'' in the sample of 
high-resolution spectroscopic Li-rich red-giant targets studied by \cite{yan2021}. Note that it is difficult to distinguish AGB and RGB stars
by asteroseismology methods because they have a similar structure. This similarity also means that if the IGW mixing were as efficient in
upper RGB stars as the RGB extra mixing it would prevent the formation of the $^{13}$C pockets for the main {\it s} process \citep{straniero1995}
in their AGB descendants that needs much slower mixing to gently inject protons into the $^4$He- and $^{12}$C-rich layers \citep{denissenkov2003b}.
The 3D hydrodynamics simulations of IGW mixing in upper RGB stars have not ruled out a possibility of such slow IGW mixing in AGB stars \citep{blouin2023}.

Unfortunately, there is not yet a model or magneto-hydrodynamics simulations of AMRI applicable to the magnetic and thermodynamic conditions
in the radiative zones of low-mass RGB stars. The simulations of AMRI by \cite{ruediger2015} assumed a uniform density distribution,
therefore they neglected the stabilizing effect of the buoyancy produced by the stable thermal stratification in
the radiative zone of an upper RGB star. To compensate for this deficiency of the simulations, we have assumed that the AMRI diffusion coefficient is
proportional to the radiative diffusivity $K$ that reduces the stabilizing buoyancy force via the exchange of heat between 
rising fluid and its surroundings. By the same reason, we have also assumed that the AMRI mixing, like the other RGB extra mixing
mechanisms, may become active only in the chemically-uniform (with $\nabla\mu\approx 0$) radiative zone above the bump luminosity. 

It is not clear yet how the empirically constrained equation  
$D_\mathrm{mix}\propto (\Omega_\mathrm{core}/\Omega_\mathrm{env})^\alpha$, with $\alpha = 2$ \citep{spada2016} or 
$\alpha = 1.5$ \citep{dumont2023}, for the AMRI diffusion coefficient is transformed into the scaling relation
$D_\mathrm{mix}\propto (L/L_\mathrm{bump})^p$ used in our parametric model of RGB extra mixing. 
The value of $p$ for the AMRI mechanism probably depends on the ratio of the core-envelope rotational
coupling time to the evolutionary time on the RGB. 
Note that in MS low-mass stars, in which the AMT appears to be more efficient than the AMT by the AMRI in their post-MS descendants \citep{dumont2023},
the coupling time is $\sim 50$ Myr \citep{denissenkov2010b}, so it is comparable to the upper-RGB evolutionary timescale.
Given that $R\propto L$ on the upper RGB,
the power may become as large as $p\approx 4$, when the RGB evolution significantly
speeds up near the RGB tip, and the AMRI angular-momentum transport may be not fast enough to decrease the resulting differential rotation
between the rapidly-rotating core and the convective envelope whose rotation is slowed down by its expansion and conservation of angular momentum. 
On the other hand, the much slower evolution of the star at the beginning of its ascent of the upper RGB may give the AMRI enough time to
reduce the ratio $(\Omega_\mathrm{core}/\Omega_\mathrm{env})$, which could result in $p < 4$. Therefore, in this work we have used
the parametric model only with the value of $p = 4/3$, that we expected to be appropriate for the models of IGW and rotational mixing 
with the modified Reimers mass-loss rate, just as a proof of concept.

When we look for observational data that could be used to substantiate or disapprove our hypothesis of Li-enrichment of RC stars
based on the AMRI mechanism of RGB extra mixing, we find the correlations of enhanced Li abundances in field and open-cluster
red giants with their rotation \citep[e.g.,][]{drake2002,ming-hao2021,tsantaki2023} to be a supporting evidence. On the other hand,
the scarcity of Li-rich HB stars in both globular clusters and open clusters with low MS turn-off masses \citep{kirby2016,sanna2020,magrini2021,tsantaki2023}
can be considered as an argument against our hypothesis, especially given that our computations predict similar enrichments of upper RGB stars
in Li by enhanced extra mixing in stars with different initial masses and metallicities (Figure \ref{fig:fig_li_diff_m}). 
However, the last observational fact could be even more difficult to explain
if the Li-enrichment of RC stars were associated with the He-core flash in their RGB progenitors because this is a universal physical process
occurring in all low-mass stars, whereas the AMRI mixing mechanism may depend on different rotational, mass-loss, and magnetic-field evolution histories of
low-mass stars which may be responsible for the observed diversity of Li-abundance distributions in HB stars in the field and in stellar clusters. 

The other possible mechanism of enhanced RGB extra mixing leading to the Li enrichment of RC stars
may be thermohaline convection assisted by radial magnetic field \citep{fraser2024}, especially given
that surprisingly strong radial magnetic fields with $B_r\sim 100$ kG have recently been measured in radiative zones of 
low-mass red giants \citep{deheuvels2023,ligang2023}. 
In the second of the cited papers it has also been reported that the core and envelope averaged angular velocities of these stars do not differ from
those measured in other red giants, which can be an indication that such magnetically-assisted mixing may be operating in all of them with
different efficiencies proportional to $B_r^2$, as predicted by \cite{fraser2024}.

Finally, we cannot exclude a possibility that our hypothesized diffusion coefficient for RGB extra mixing increasing with the luminosity towards
the RGB tip may result from an intricate combination of different mixing processes manifesting themselves with different efficiencies at
different evolutionary phases.

\section*{Data availability}

The data underlying this article will be shared on reasonable request to the corresponding author.

\section*{Acknowledgements}

SB is a Banting Postdoctoral Fellow and a CITA National Fellow, supported by the
Natural Sciences and Engineering Research Council of Canada
(NSERC). FH acknowledges funding through an NSERC Discovery Grant. PRW acknowledges funding through NSF grants 1814181
and 2032010. FH and PRW have been supported through NSF award
PHY-1430152 (JINA Center for the Evolution of the Elements).
The computations and data analysis were carried on the Astrohub online virtual research environment 
(\href{https://astrohub.uvic.ca}{https://astrohub.uvic.ca})
developed and operated by the Computational Stellar Astrophysics
group (\href{https://csa.phys.uvic.ca}{https://csa.phys.uvic.ca}) at the University of Victoria
and hosted on the Compute Canada Arbutus Cloud at the University
of Victoria.




\bibliographystyle{mnras}
\bibliography{lirich.bib}





\bsp	
\label{lastpage}
\end{document}